\documentclass[journal, twoside, final]{IEEEtran}
\pdfoutput=1
\usepackage{cite}
\usepackage[pdftex]{graphicx}
\usepackage{amssymb, amsmath, amsthm}
\usepackage{amsfonts}
\usepackage{multirow}
\usepackage{mathrsfs}
\usepackage{color}
\usepackage{url}
\usepackage{flushend}
\usepackage{setspace}
\usepackage{placeins}
\usepackage{wrapfig}
\usepackage{booktabs, multicol, multirow}
\usepackage{slashbox}

\usepackage{makecell}
\allowdisplaybreaks

\newtheorem{special cases}{Special Cases}

\newcommand{\ie}{{\sl i.e.}}
\newcommand{\eg}{{\sl e.g.}}

\begin{document}
\title{Low-Cost Anti-Copying $2$D Barcode by Exploiting Channel Noise Characteristics}

\author{Ning~Xie,~\IEEEmembership{Senior Member,~IEEE}, Qiqi~Zhang, Ji~Hu, Gang~Luo, and Changsheng~Chen,~\IEEEmembership{Member,~IEEE}
\vspace{-1cm}
\thanks{%
The authors are with the Guangdong Key Laboratory of Intelligent Information Processing, College of Information Engineering, Shenzhen University, Shenzhen, 518060, China (e-mail: ningxie@szu.edu.cn; cschen@szu.edu.cn).
}
}

\maketitle
\begin{abstract}
%
%
In this paper, for overcoming the drawbacks of the prior approaches, such as low generality, high cost, and high overhead, we propose a Low-Cost Anti-Copying (LCAC) $2$D barcode by exploiting the difference between the noise characteristics of legal and illegal channels.
An embedding strategy is proposed, and for a variant of it, we also make the corresponding analysis.
For accurately evaluating the performance of our approach, a theoretical model of the noise in an illegal channel is established by using a generalized Gaussian distribution.
By comparing with the experimental results based on various printers, scanners, and a mobile phone, it can be found that the sample histogram and curve fitting of the theoretical model match well, so it can be concluded that the theoretical model works well.
For evaluating the security of the proposed LCAC code, besides the direct-copying (DC) attack, the improved version, which is the synthesized-copying (SC) attack, is also considered in this paper.
Based on the theoretical model, we build a prediction function to optimize the parameters of our approach.
The parameters optimization incorporates the covertness requirement, the robustness requirement and a tradeoff between the production cost and the cost of illegally-copying attacks together.
The experimental results show that the proposed LCAC code with two printers and two scanners can detect the DC attack effectively and resist the SC attack up to the access of $14$ legal copies.
\end{abstract}


\IEEEpeerreviewmaketitle

\section{Introduction}
Two-Dimensional ($2$D) barcodes are widely used in various applications because of their advantages of simple and low cost.
In addition, one attractive feature in $2$D barcodes is capable of providing significantly higher information capacity than that in $1$D barcodes \cite{8013082, 6894205}.
A $2$D barcode pattern named Quick Response (QR) code has been popularly used in our daily life.
For example, a QR code can be employed as the information entrance of an advertisement, the information carrier for a mobile payment transaction, and a product authentication for tracking and anti-counterfeiting, etc.

Recently, the security of $2$D barcodes has received extensive attention due to the following three major security risks \cite{6952257,7472057,7349185,8794824,7902126,8630792}.
First, various types of illegal information, \eg, Trojan virus and phishing websites, are encoded in a normal $2$D barcode.
It is challenging to detect illegal information before the barcode is decoded \cite{Krombholz2014QR}.
Second, a $2$D barcode can be illegally tampered by a replacement attack that covers the original barcode by an illegal one.
Under such attacks, some important information, \eg, the payee of a mobile payment transaction, can be tampered and it results in may cause some economic loss \cite{Zhu2016Secure}.
Third, a $2$D barcode can be illegally replicated to fake a unique identifier in a tracking system for the anti-counterfeiting application.

The first two security risks have been effectively overcome.
For example, for the first security risk, an anti-virus and anti-phishing recognition mechanism was used before the receiver of a $2$D barcode executes the decoded information \cite{Vidas2013QRishing}, while for the second security risk, the digital signature algorithms can be used to check the authenticity and integrity of the contents in a $2$D barcode \cite{Vill2006Multilevel}.
However, the third security risk (illegal copying) is more challenging as compared with the other two risks, since a $2$D barcode can be easily replicated with an off-the-shelf photocopier.
An illegal copying $2$D barcode not only leads to large economic and reputational loss for the authorized manufacturer but also limits the application of $2$D barcodes as an anti-counterfeiting technique.
Thus, this paper focuses on the problem of illegal copying.


In the literature, some approaches have been proposed to overcome the security risk of illegal copying but accompanying with some limitations. Now, we briefly introduce them as follow.

1) \emph{Special Printing Materials or Techniques}. This approach exploits the special features of printing materials or techniques, which cannot be reproduced on purpose, to counter the attack of illegal copying. For example, a polymerized liquid crystal material \cite{Gremaud2015Identification} with unique optical characteristics can be used to print an anti-copying $2$D barcode. Some red, green and blue light-emitting nano-particles \cite{You2016Three} can be used to construct 3-dimensional ($3$D) QR codes that cannot be copied by ordinary technologies. Special halftone printing technology \cite{Maehara2014Watermark} can generate $2$D barcodes that are invisible under visible light. However, this approach not only increases the production cost but also reduces the universal applicability of a $2$D barcode, which hinders its promotion in extensive applications.

2) \emph{Physical Unclonable Function (PUF)}. The PUF is an unclonable response function which inputs a stimulus to a physical entity and then outputs a unique feature according to the internal physical structure, \eg, a unique texture of printing paper \cite{Voloshynovskiy2016Physical}. In recent years, researchers have found that it is possible for a mobile imaging device under a semi-controlled condition to acquire images of paper and to extract microscopic textural features for constructing the PUF \cite{Wong2016Counterfeit}. The PUF, which acts as a digital signature of each printing substrate, is stored in an online database to facilitate the verification of textural features extracted from a query document. This approach has a limitation. The authentication is performed over an online database, where the scale of the database has been greatly restricted. The scale of the database applied is often not sufficiently large to have extensive universality. When the scale of the database is expanded, the accuracy of its authentication will be reduced.

3) \emph{Anti-copying Pattern}. Some patterns with detailed features, such as high-density black or white blocks, can be used to prevent illegal copying \cite{Tkachenko2016Centrality}. Similarly, the following patterns also can be used for anti-copying, \eg, the black-and-white texture pattern with a grating structure \cite{8819920} and color anti-copying pattern which contains the information of four channels of CMYK \cite{Zhao2011Frequency}. This approach requires that the legal receiver equips a capturing device with a high resolution, which apparently increases the implementation cost of the receiver and is impractical for a low-cost mobile phone.

4) \emph{Digital Watermarking}. The digital watermarking technology (DWT) can embed certain privacy information in a $2$D barcode so as to protect its content authenticity \cite{Maehara2014Watermark}. Semi-fragile watermark is an important branch of DWT \cite{Malarchelvi2013ASI}, which can resist distortion or tamper with low intensity and can detect distortion or falsification of various types of images with high intensity \cite{Xie2015Anti}. However, to our best knowledge, there is no public report that a digital watermarking technique has been used against illegally-copying attacks.

In summary, the existing anti-copying approaches have the drawbacks of low generality for special material, high cost for high-resolution equipment and high overhead for online database required. In this paper, we focus on the print-capture channel where a message is transmitted using a printed medium and is retrieved by a mobile imaging device. We propose a low-cost anti-copying (LCAC) $2$D barcode by exploiting the difference between the noise characteristics of legal and illegal channels. At the same time, we propose some possible transformations of the $2$D barcode, and give corresponding covertness and robustness analysis. For accurately evaluating the performance of the proposed LCAC code, a theoretical model of the noise in an illegal channel is established by using a generalized Gaussian distribution. At last, based on the theoretical model, we built a prediction function to optimize the parameters of the proposed LCAC $2$D barcode in order to increase the cost of copying attack and achieve a better anti-copying effect.

The key contributions of this work can be summarized as follows.
\begin{enumerate}
  \item We propose a low-cost anti-copying (LCAC) $2$D barcode on the basis of the considered $2$D barcode, which exploits the difference between the legal and illegal channels. In the proposed LCAC code, the sender of a $2$D barcode embeds an authentication message into a source message to realize the anti-copying purpose. Two embedding strategies are proposed and analyzed.
  \item For accurately evaluating the performance of the proposed LCAC code, a theoretical model of an illegal channel based on various printers and scanners is established by using a generalized Gaussian distribution. By comparing with the actual experimental results, the theoretical model works well.
  \item For improving the security of the proposed LCAC code, besides the direct-copying attack, an improved version which is the synthesized-copying attack, is also considered in this paper. Based on the aforementioned theoretical model, we built a prediction function to optimize the parameters of the proposed LCAC code. The parameters optimization incorporates the covertness requirement, the robustness requirement and a tradeoff between the production cost and the cost of illegally-copying attacks together. The experimental results show that our approach has a good ability to prevent illegal copying.
\end{enumerate}


\section{Background of Considered $2$D Barcode and System Model}
\subsection{Background of Considered $2$D Barcode}
\begin{figure*}[!t]
\centering
\includegraphics[width=6.5in]{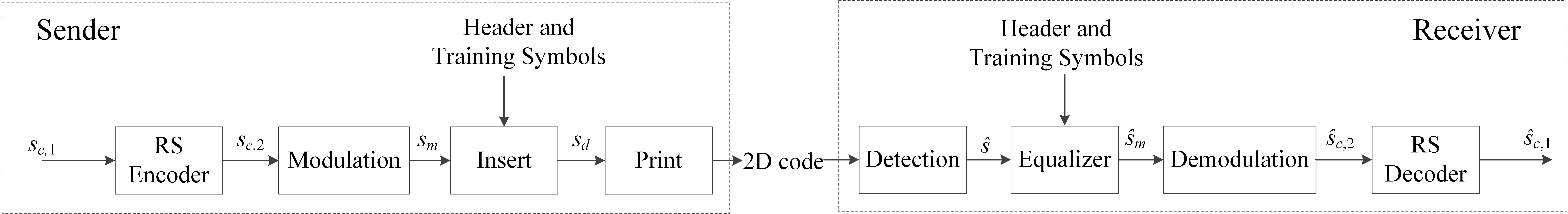}
\caption{Block diagram of the considered $2$D barcode.}
\label{Figure1}
\vspace{-0.5cm}
\end{figure*}

Without loss of generality, this paper considers $M$-order multilevel $2$D barcodes \cite{Vill2006Multilevel}, where $M$ is the modulation order and $M\ge 2$. The block diagram of the generic $2$D barcode is illustrated in Fig. \ref{Figure1}, as shown at the top of the next page.
As shown in the sender of Fig. \ref{Figure1}, ${{s}_{c,1}}$ denotes a source message with length ${{L}_{c}}=Nk$, where $N$ is the number of blocks and $k$ is the length per block. The ${{s}_{c,1}}$ is encoded via Reed-Solomon (RS) codes to obtain the coder output ${{s}_{c,2}}$. The output length of RS codes per block is denoted as n and its error correction capability is $t=\left( n-k \right)/2$. Thus, the length of ${{s}_{c,2}}$ is ${{L}_{s}}=Nn$.

Then, through a pulse amplitude modulation (PAM) with order $M$\cite{Sakib2013A}, we obtain a modulated signal ${{s}_{m}}$. The modulate block transforms a bit stream into the corresponding gray-scale values. The case of $M=2$ is very popular in the practical application of $2$D barcodes; however, the cases of $M\ge 4$ is becoming a new research trend due to its high capacity \cite{Vill2006Multilevel}. Thus, this paper focuses on the case of $M=4$. Following \cite{zhang2019accurate}, the constellation points $x$ are set as $x\in \left\{ 40,100,160,220 \right\}$, that is, ${{x}_{1}}=40,$ ${{x}_{2}}=100,$ ${{x}_{3}}=160,$ ${{x}_{4}}=220$. Note that the ideas of this paper can be straightforwardly extended to other cases, \eg, $M=2$ and $M=8$.

\begin{figure}[!t]
\centering
\begin{minipage}[t]{0.48\linewidth}
\includegraphics[width=.95\textwidth]{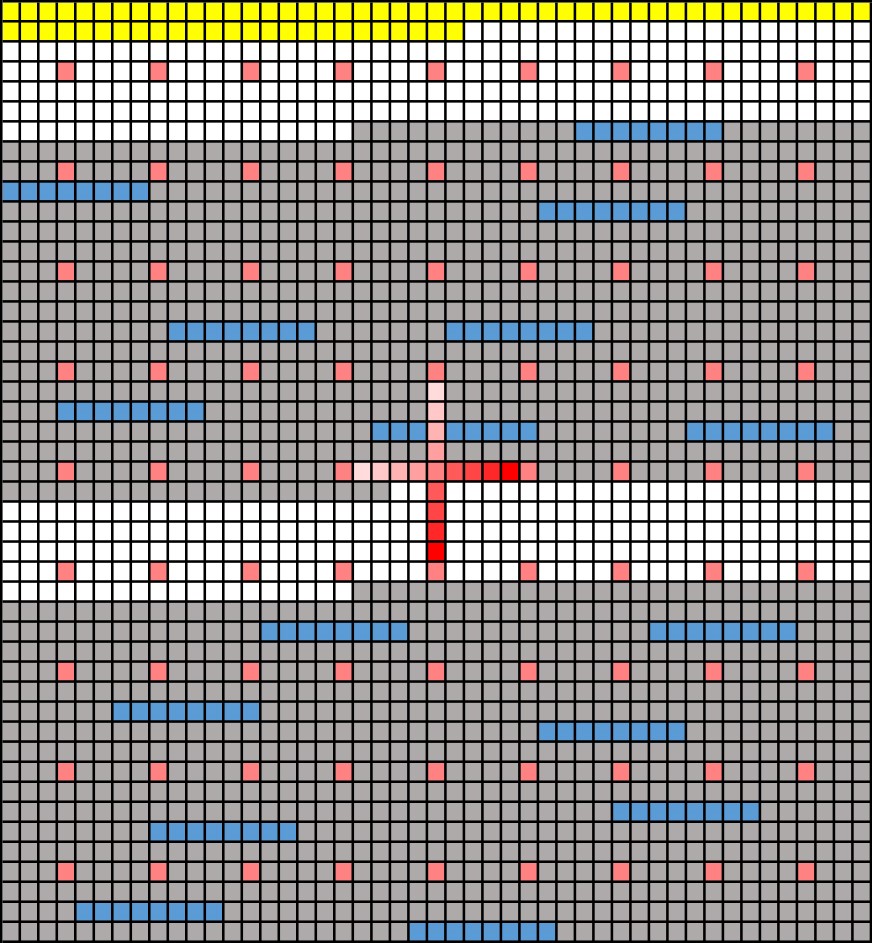}
\footnotesize\centerline{(a)}
\end{minipage}
\begin{minipage}[t]{0.483\linewidth}
\centering
\includegraphics[width=.95\textwidth]{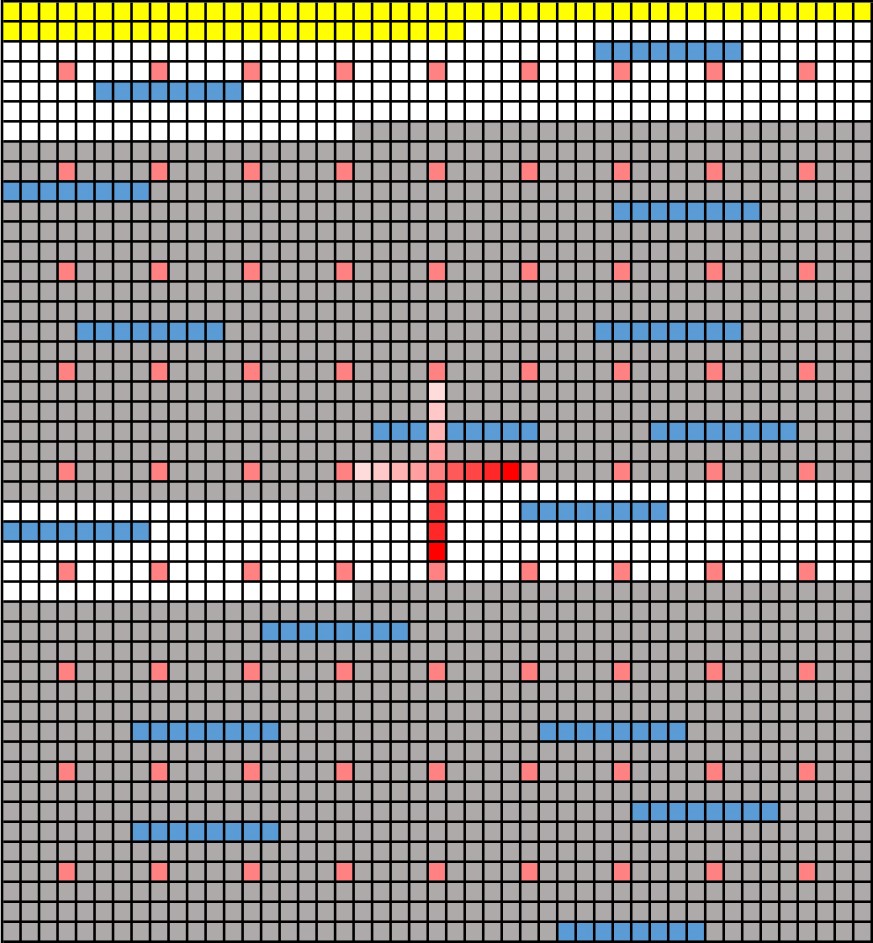}
\footnotesize\centerline{(b)}
\end{minipage}
\caption{Structure diagrams of LCAC codes, where the number of RS code blocks is 2 ($N=2$), yellow and red modules represent the header and training symbols, respectively. The two diagrams are based on two different embedding strategies: (a) Strategy $1$; (b) Strategy $2$.
In Strategy $1$, the authentication message is randomly embedded into the redundant bits of the source message, whereas, in Strategy $2$, the authentication message is randomly embedded into the entire bits of the source message.}
\label{Figure2}
\vspace{-0.5cm}
\end{figure}


After inserting the header and training symbols into the modulated signal ${{s}_{m}}$, the original version of a $2$D barcode, ${{s}_{d}}$, is generated. The total length of header and training symbols is denoted as ${{L}_{h}}$ and the final length of a $2$D barcode is denoted as ${{L}_{t}}={{L}_{s}}+{{L}_{h}}$, where the value of ${{L}_{t}}$ should be an integer after taking a square root to keep the $2$D barcode a square structure. The header symbols have two functions: first, it stores additional information, \eg, format and version of a considered $2$D barcode; second, its length is adjustable to ensure that the value of ${{L}_{t}}$ satisfies the length requirement in a considered $2$D barcode \cite{zhang2019accurate}.

In Fig. \ref{Figure2}, the white and grey modules represent the modulated symbols of information bits and redundant bits of the source message, respectively. The yellow and red modules represent the locations of header symbols and training symbols, respectively. The intensity of red illustrates the gray level of the training symbols. Moreover, two types of training symbols are considered. The first type is used to estimate the spatial distortion, which is set over an entire $2$D barcode as uniform as possible; the second type is used to estimate the post-processing distortion, which is set at the center area of a $2$D barcode \cite{zhang2019accurate}. Following \cite{zhang2019accurate}, the gray value of the first type of training symbols is set to $130$, while those of the second type of training symbols are set to $\left\{ 30,50,70,100,160,180,200,220 \right\}$, as shown the red cross of Fig. \ref{Figure2}. Note that, as the last step of encoding a $2$D barcode, a finder pattern should be attached to facilitate the detection of a $2$D barcode reader \cite{rungraungsilp2012data}; however, without causing any confusion, the finder pattern is omitted in Fig. \ref{Figure1} for conciseness.

In Fig. \ref{Figure1}, the first block of a receiver is the detection block which scans the printed $2$D barcode with a mobile phone camera or an optical scanner. The detection block is to locate and to segment the captured version of a $2$D barcode by using the finder pattern. Then, the detection block quantizes the mean intensity within each partitioned area to obtain a gray-scale signal ${{\hat{s}}_{d}}$. The ${{\hat{s}}_{d}}$ is fed into an equalizer block which compensates the channel distortion by using the training symbols described above \cite{zhang2019accurate}. Specifically, the equalizer block trains a fitting function to reflect the channel distortion by comparing the gray-scale values of the scanned training symbols with those of the considered ones. The fitting function can be described by a sigmoid function in practical situations \cite{zhang2019accurate}. When the fitting function is trained, an inverse fitting function is further established to correct the distortions in ${{\hat{s}}_{d}}$. Then, ${{\hat{s}}_{m}}$ is extracted by removing the header and training symbols in ${{\hat{s}}_{d}}$. Next, ${{\hat{s}}_{m}}$ is demodulated and decoded to obtain ${{\hat{s}}_{c,2}}$ and ${{\hat{s}}_{c,1}}$, respectively.

\subsection{System Model}
\begin{figure}[!t]
\centering
\includegraphics[width=3.5in]{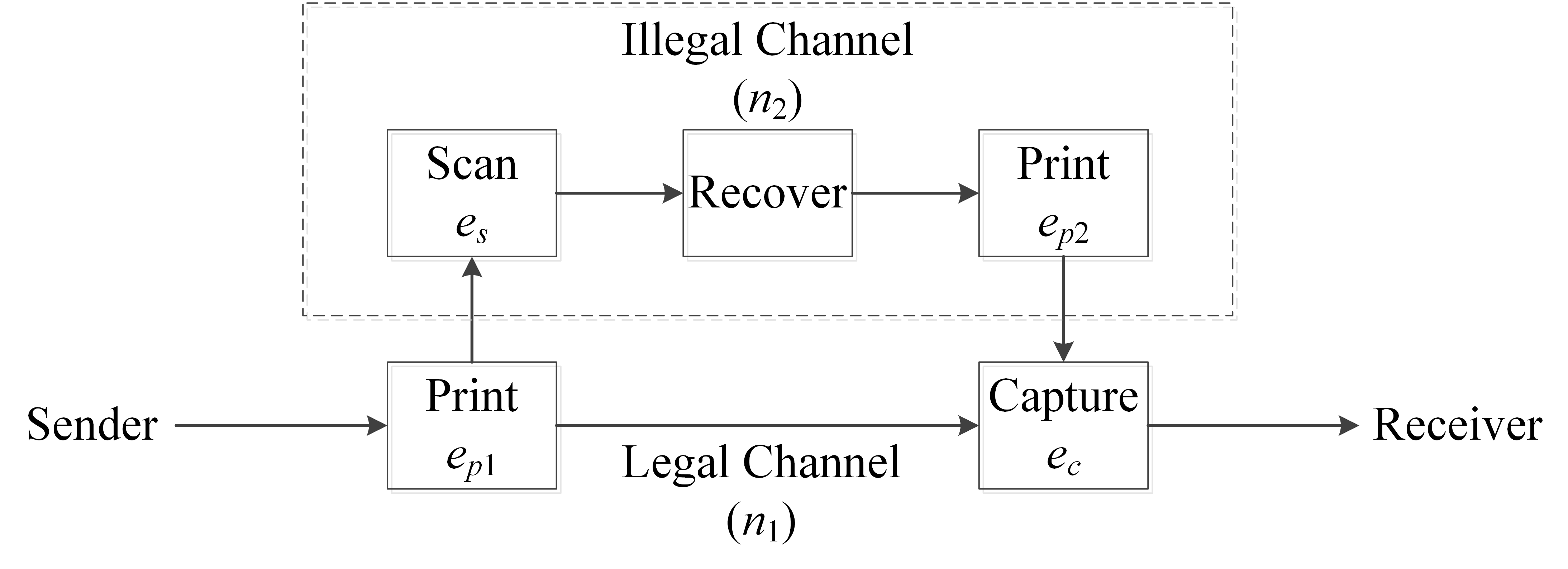}
\caption{The system model of a $2$D barcode with two possible channels, \ie, legal channel and illegal channel.}
\label{Figure3}
\vspace{-0.5cm}
\end{figure}
As shown in Fig. \ref{Figure3}, we consider the system model of a $2$D barcode by two different channels, \ie, a legal channel and an illegal channel. In the legal channel, as shown by the lower branch of Fig. \ref{Figure3}, a legal $2$D barcode is received through only a print-and-capture process, whereas in the illegal channel as enclosed by the dashed box of Fig. \ref{Figure3}, an illegal $2$D barcode is received through a print-scan-print-and-capture process, which is denoted as a double print $\&$ scan (DPS) process. Intuitively, the distortion and noise in an illegal channel are more serious than those in a legal channel.
Specifically, the total noise in a legal channel can be modeled as
\begin{equation}
{{e}_{1}}={{e}_{{{p}_{1}}}} \oplus {{e}_{c}},
\label{equ1}
\end{equation}
where  ${{e}_{{{p}_{1}}}}$ and  ${{e}_{c}}$ represent the noise components of the first printing process and the legal detecting process, respectively, $'\oplus'$ represents the interaction of noise in different stages. The more common relationships are additive noise and multiplicative noise. The total noise in an illegal channel is written as
\begin{equation}
{{e}_{2}}={{e}_{{{p}_{1}}}} \oplus {{e}_{s}} \oplus {{e}_{{{p}_{2}}}} \oplus {{e}_{c}},
\label{equ2}
\end{equation}
where  ${{e}_{s}}$ and  ${{e}_{{{p}_{2}}}}$ denote the noise components of the illegal scanning process and the second printing process, respectively.
\begin{figure}[!t]
\centering
\begin{minipage}[t]{0.18\linewidth}
\includegraphics[width=.95\linewidth]{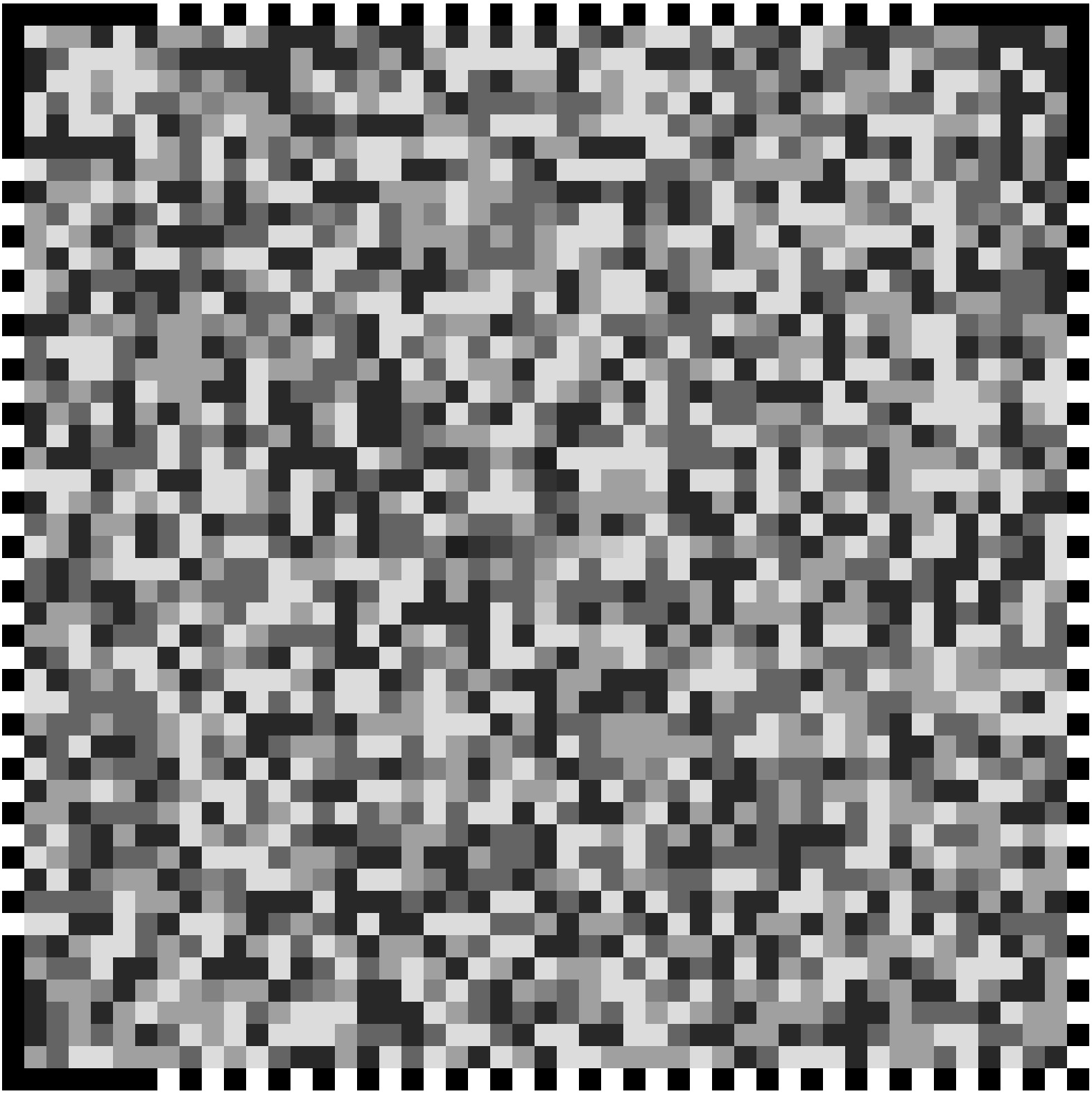}
\footnotesize\centerline{(a)}
\end{minipage}
\begin{minipage}[t]{0.18\linewidth}
\includegraphics[width=.95\linewidth]{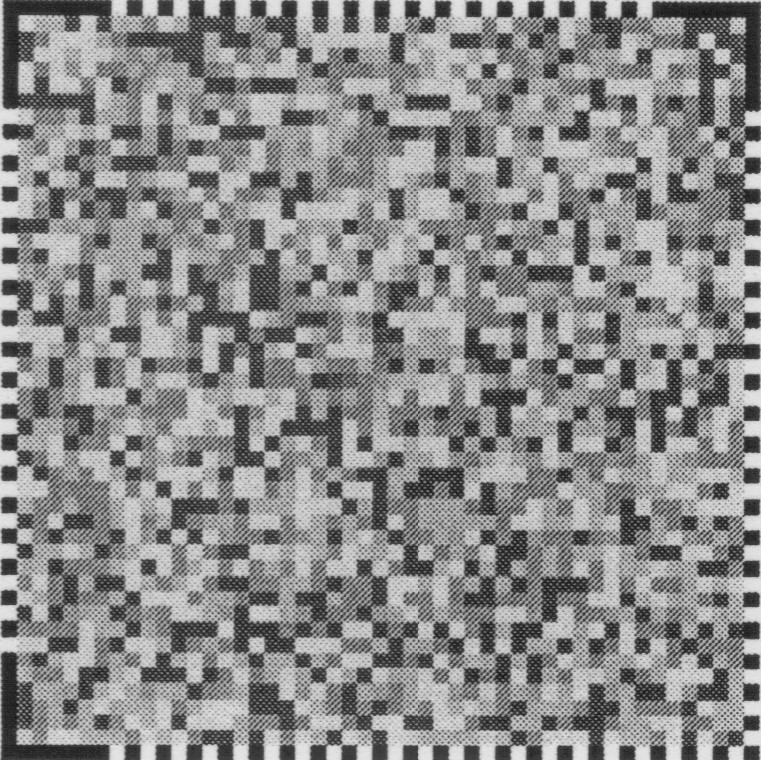}
\footnotesize\centerline{(b)}
\end{minipage}
\begin{minipage}[t]{0.18\linewidth}
\includegraphics[width=.95\linewidth]{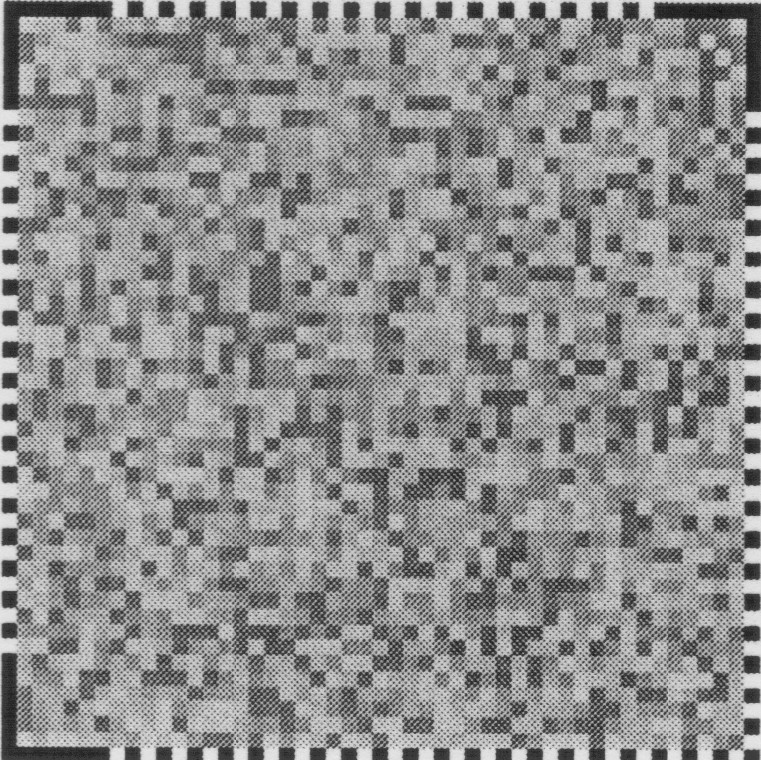}
\footnotesize\centerline{(c)}
\end{minipage}
\begin{minipage}[t]{0.18\linewidth}
\includegraphics[width=.95\linewidth]{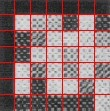}
\footnotesize\centerline{(d)}
\end{minipage}
\begin{minipage}[t]{0.18\linewidth}
\includegraphics[width=.95\linewidth]{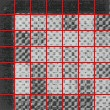}
\footnotesize\centerline{(e)}
\end{minipage}
\caption{Examples of a $2$D barcode: (a) original $2$D barcode; (b) captured legal $2$D barcode; (c) captured illegally copied $2$D barcode; (d) enlarged top-left region of a legal $2$D barcode; (e) enlarged top-left region of an illegally copied $2$D barcode.}
\label{Figure4}
\vspace{-0.5cm}
\end{figure}

By considering the models of printing and scanning processes in \cite{zhang2019accurate}, it is easy to conclude that $\sigma _{{{e}_{1}}}^{2}<\sigma _{{{e}_{2}}}^{2}$ since more processes of printing or scanning introduce more noise.
This conclusion is also demonstrated through an example shown in Fig. \ref{Figure4}.
By comparing the details of Fig. \ref{Figure4}(a) with those of Fig. \ref{Figure4}(b) and Fig. \ref{Figure4}(c), the noise variance of the legal $2$D barcode is slightly larger than that of  the original barcode, whereas the noise variance of the illegally copied one is much larger than those of both the legal $2$D barcode and the original barcode.
Although it is possible to distinguish the illegally copied $2$D barcode by detecting various characteristics of two channels \cite{Tkachenko2016Centrality,Ho2014Document}, it is challenging for a low-cost mobile terminal, \eg, a mobile phone, to finish this task due to the following two reasons. First, it is difficult to set an appropriate threshold to distinguish two types of $2$D barcodes, even for the enlarged subgraph as shown in Fig. \ref{Figure4}(d) and Fig. \ref{Figure4}(e), since there is no prior information for two types of channels. Second, if the mobile phone performs an authentication with the online database, which not only causes large network overhead but also introduces additional security risks due to the frequent sensitive data exposure during wireless communication.
The basic idea of our approach is to embed an authentication message into a source message, where the decoded bit-error-rates (BER) of the authentication message is sensitive to the noise variance.
In the receiver, we set an appropriate threshold for decoded BER of authentication message to detect whether the received $2$D barcode is an illegally copied version of the legal $2$D barcode or not, which will be described in the next section.
Then, we fine-tune the performance of our approach by optimizing the embedding parameters of the authentication message, which will be described in Section VI.

\section{The Proposed Low-cost Anti-Copying Code}
\subsection{Description for the Sender}
\begin{figure}[!t]
\centering
\includegraphics[width=3.5in]{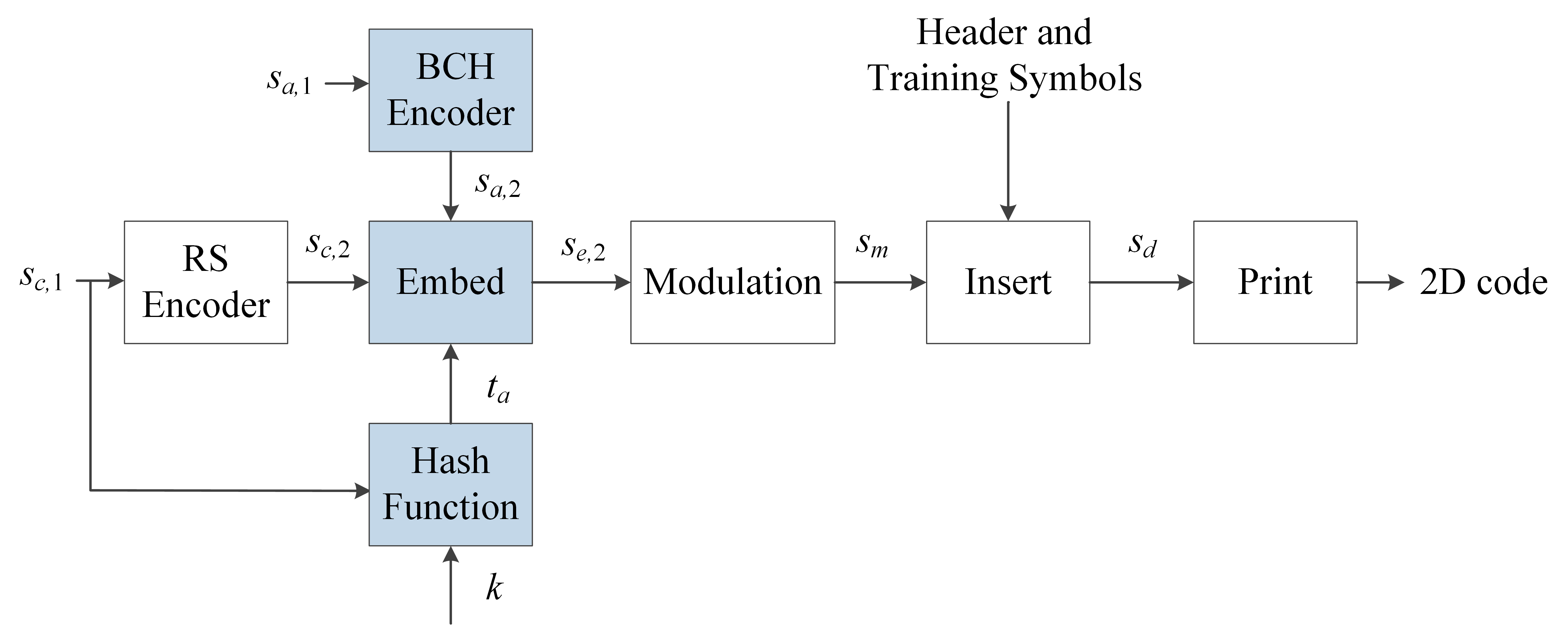}
\caption{Block diagram of the sender in the proposed LCAC code.}
\label{Figure5}
\end{figure}
As shown in Fig. \ref{Figure5}, we generate an authentication message  ${{s}_{a,1}}$ for anti-copying purpose. The ${{s}_{a,1}}$ is encoded via the Bose-Chaudhuri-Hocquenghem (BCH) codes to obtain the coder output ${{s}_{a,2}}$ for improving its robustness. The lengths of ${{s}_{a,1}}$ and ${{s}_{a,2}}$ are denoted as ${{k}_{a}}$ and ${{n}_{a}}$ respectively, and the error correction capability is ${{t}_{a}}$.

The key block of the sender in the proposed LCAC code is the embed block which replaces certain bits of the source message by those of the authentication message. Specifically, if the bit of the source message is different from that of the authentication message, this bit is modified from $0$ to $1$ or $1$ to $0$; otherwise, this bit is kept unchanged. Apparently, the embedding operation sacrifices the robustness of the source message, which also can be denoted as the covertness of the authentication message. There are two aspects of the covertness requirement in our LCAC code. First, the presence of the authentication message should not be easily detectable by the illegal receiver. Second, it should not have a noticeable effect on the receivers' ability to recover the source message. Moreover, covert authentication in an LCAC code may be used together with other security techniques in the conventional approaches to produce a more secure $2$D barcode. In this paper, the covertness performance is analyzed through the error probability of demodulation and decoding for the source message. The values of ${{k}_{a}}$ and ${{n}_{a}}$, the parameters of the proposed LCAC code, should be optimized by jointly considering the covertness and robustness of authentication message, which will be analyzed in Section VI.

Besides the embedding length ${{n}_{a}}$, The embedding locations should be carefully designed as well. In this paper, we consider two embedding strategies to define different embedding locations. In the first strategy, as shown in Fig. \ref{Figure2}(a), which is short as Strategy $1$ for simplicity, ${{s}_{a,2}}$ is randomly embedded into the redundant bits of ${{s}_{c,2}}$. In the second strategy, as shown in Fig. \ref{Figure2}(b), which is short as Strategy $2$, ${{s}_{a,2}}$ is randomly embedded into the entire bits of ${{s}_{c,2}}$. In the sender of an LCAC code, the specific embedding locations are defined through a one-way, collision-resistant hash function with the source message and the secret key $k$, expressed as
\begin{equation}
{{t}_{a}}=g\left( {{s}_{c,1}},k \right),
\label{equ3}
\end{equation}
where the hash function $g\left( \cdot  \right)$ is robust against input error for generating the random locations.
The secret key $k$ is generated and allocated by the sender.
For higher security, $k$ is different for different source messages ${{s}_{c,1}}$, which means that different keys are generated for different products. Before each verification attempt, the legal receiver sends an authentication request to the sender, and the sender feedbacks $k$ and ${{s}_{a,1}}$ to the receiver via a secure way, \eg, encryption.
Note that we can use some advanced key agreement protocols \cite{WANG20104052, NIU20111986} to achieve the allocation of secret keys for further improving the security of the considered 2D barcode.

Note that both $k$ and ${{s}_{a,1}}$ are only secret information required in the LCAC codes and the exchange of secret information occurs only once, thus the overhead for anti-copying in the LCAC is very low. An alternative solution is to allocate both $k$ and ${{s}_{a,1}}$ when the authentication program is installed into a mobile phone, which avoids the exchange of secret information.
Note that, before a legal receiver accepts a $2$D barcode, the authentication message should be treated as a noise signal as well since the legal receiver does not know whether the authentication message exists or not in the received signal.

For two embedding strategies, we have the following observations:

\emph{Observation $1$}: Strategy $1$ has better covertness performance than Strategy $2$ since the impact of errors occurred in the redundant bits to the decoding performance of source message is smaller than that in the information bits according to the coding structure of RS codes.

\emph{Observation $2$}: Strategy $2$ has better robustness performance than Strategy $1$ since Strategy $2$ has larger embedding range, which spreads the authentication message into a larger area and lowers the error probability caused by some local distortion, \eg, a part of the $2$D barcode is shaded.

\emph{Observation $3$}: Strategy $2$ has better security performance than Strategy $1$ since each bit of the authentication message is embedded in a larger space, which intuitively increases the uncertainty of detecting the authentication message by an adversary. The first two observations are verified through experimental results in  APPENDIX A.

The two embedding strategies in LCAC codes are illustrated in Fig. \ref{Figure2} as two toy examples, where the case of $N=2$ is considered. The blue modules represent the modulated symbols of the authentication message. Note that, as the last step of encoding a $2$D barcode, a finder pattern should be attached to facilitate the detection of a $2$D barcode reader \cite{rungraungsilp2012data}; however, without causing any confusion, the finder pattern is omitted in Fig. \ref{Figure2} for conciseness.

After the embed block, we obtain a signal with embedding message ${{s}_{e,2}}$. Then, through a pulse amplitude modulation (PAM) with order $M$ \cite{Sakib2013A} to yield a modulated signal ${{s}_{m}}$. After inserting the header and training symbols into the modulated signal ${{s}_{m}}$, the original version of a $2$D barcode, ${{s}_{d}}$, is generated.

\subsection{Description for the Receiver}
\begin{figure}[!t]
\centering
\includegraphics[width=3.5in]{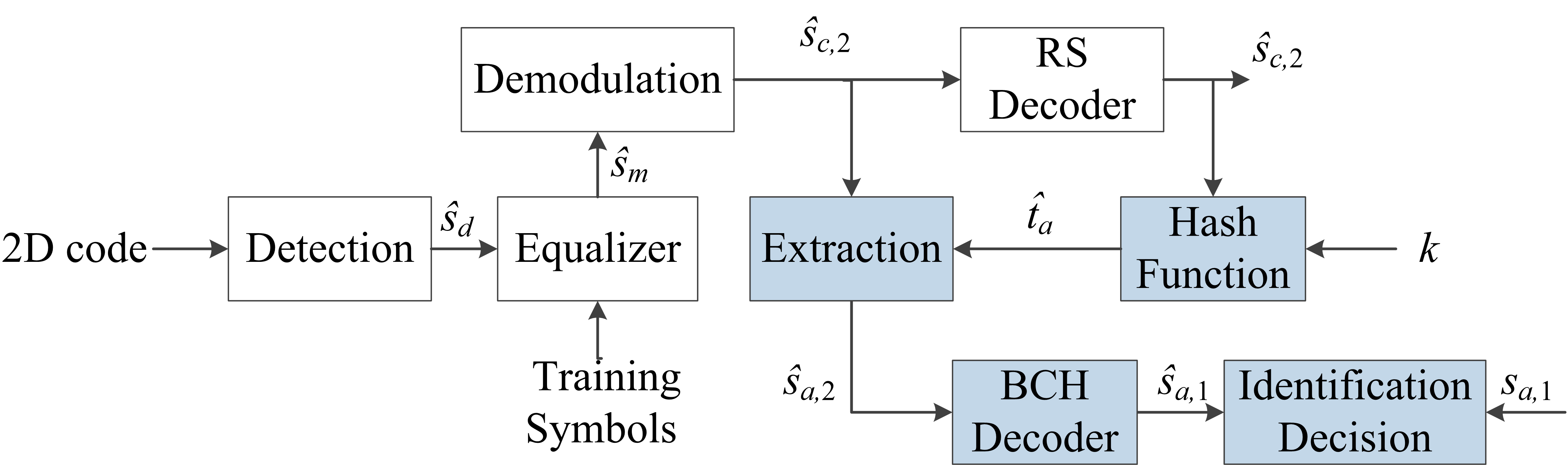}
\caption{Block diagram of the receiver in the proposed LCAC code.}
\label{Figure6}
\end{figure}
When a printed $2$D barcode is captured by a receiver, a series of standard operations should be proceeded to extract the source message. The block diagram of the receiver in the proposed LCAC Code is illustrated in Fig. \ref{Figure6}. Similar to Fig. \ref{Figure1}, the blue blocks represent the additional one for anti-copying but the remaining blocks are the same as those of considered $2$D barcodes. The first block of a receiver is the detection block which scans the printed $2$D barcode with a mobile phone camera or even optical scanner. Fig. \ref{Figure4}(d) and Fig. \ref{Figure4}(e) illustrate the results of the detection block in a legal $2$D barcode and an illegally copied $2$D barcode, respectively. Then, the detection block quantizes the mean intensity within each partitioned area to obtain a gray-scale signal ${{\hat{s}}_{d}}$.

In the verification process, the receiver first generates the estimated embedding locations ${{\hat{t}}_{a}}$ using (\ref{equ3}) with the secret key $k$. And, the ${{\hat{t}}_{a}}$ can be generated without error (${{\hat{t}}_{a}}={{t}_{a}}$) even when ${{\hat{s}}_{c,1}}$ contains some error since $g\left( \cdot  \right)$ is robust against input error, \eg, robust hash functions \cite{Swaminathan2006Robust,Fridrich2000Robust}. According to the location specified by ${{\hat{t}}_{a}}$, the receiver extracts ${{\hat{s}}_{a,2}}$ from ${{\hat{s}}_{c,2}}$ through the extraction block and decodes it via a BCH Decoder to obtain ${{\hat{s}}_{a,1}}$. By comparing the values of ${{\hat{s}}_{a,1}}$ and ${{s}_{a,1}}$, the receiver makes a final authentication decision. For example, if the number of different bits between ${{\hat{s}}_{a,1}}$ and ${{s}_{a,1}}$ is beyond a predetermined threshold $\delta$, the questioned $2$D barcode is judged as an illegal one; otherwise, it is a legal one. The specific value of threshold $\delta$ is determined by exploiting the characteristics of the illegal channel.
The model analysis of illegal channel and parameter optimization of the proposed LCAC code are presented in the following two sections, Section IV, Section V, respectively.

\section{Theoretical Modeling of an Illegal Channel}
This section describes the modeling process of an illegal channel based on various printers and scanners. A typical Print \& Scan channel introduces several types of distortions, \eg, intensity variation, scaling, rotation, low-pass filtering, aliasing, and noise. The single print \& scan (SPS) process has been modeled and analyzed \cite{Vill2006Multilevel}. In this work, a series of experiments have been conducted to model an illegal channel in a DPS process.
The devices used and the corresponding parameters are listed in Tab. \ref{Table1}, where two printers, two scanners, and one mobile phone are chosen from various manufacturers.
Here, the remaining parameters of these devices are set as their default values.

\begin{table}[!t]
\centering
\caption{Description of printers, scanners and a mobile phone}
\footnotesize
\begin{tabular}{|c|c|c|}
\hline  
Name&Model&Resolution\\
\hline  
Laser Printer 1 $({{\mathsf{P}}_{1}})$&HP LaserJet P1108&1200 DPI\\
\hline 
Laser Printer 2 $({{\mathsf{P}}_{2}})$&FUJI P355D&600 DPI\\
\hline
CCD Scanner 1 $({{\mathsf{S}}_{1}})$&BENQ K810&1200 DPI\\
\hline
CCD Scanner 2 $({{\mathsf{S}}_{2}})$&EPSON V330&600 DPI\\
\hline
Mobile Phone $({{\mathsf{M}}})$& HONOR V20&48 MP\\
\hline
\end{tabular}
\label{Table1}
\vspace{-0.5cm}
\end{table}

\begin{figure}[!t]
\centering
\includegraphics[width=3.5in]{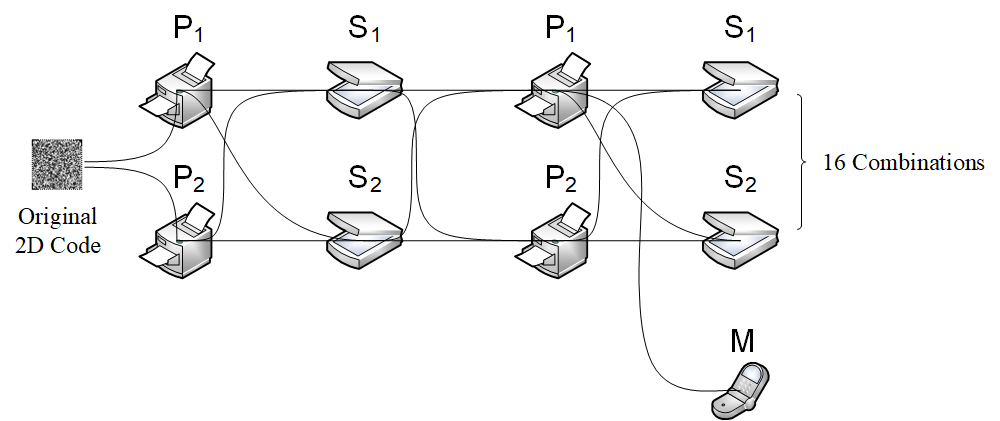}
\caption{Combinations of $2$ printers, $2$ scanners and $1$ mobile phone to emulate a DPS process.}
\label{Figure7}
\vspace{-0.5cm}
\end{figure}

Without loss of generality, the source and authentication messages are uniformly generated as two random sequences, thus, the number of symbols in different constellations are roughly equal. As illustrated in Fig. \ref{Figure7}, a total of 16 combinations is available to emulate a DPS process with $2$ printers and $2$ scanners in Tab. \ref{Table1} plus one combination of ${{\mathsf{P}}_{\mathrm{1}}}\mathrm{-}{{\mathsf{S}}_{\mathrm{1}}}\mathrm{-}{{\mathsf{P}}_{\mathrm{1}}}\mathrm{-}\mathsf{M}$.

Similar to the findings in \cite{Vill2006Multilevel}, by observing the experimental results, the intensity variation in the barcode over a DPS channel can be modeled with a generalized Gaussian distribution (GGD). For a GGD random variable (RV), \ie, $X\sim \mathcal{GGD}\left( \mu ,\sigma^2 ,\gamma  \right)$ there are three parameters, including the mean $\mu $, the variance $\sigma^2 $, and the shape factor $\gamma $. According to \cite{Nadarajah2005A}, the PDF and CDF of $X$ can respectively be given as
\begin{equation}
{{f}_{X}}\left( x \right)=\frac{\gamma \eta \left( \sigma ,\gamma  \right)}{2\Gamma \left( 1/\gamma  \right)}\exp \big[ -{{\left( \eta \left( \sigma ,\gamma  \right)\left| x-\mu  \right| \right)}^{\gamma }} \big],
\label{equ4}
\end{equation}
and
\begin{equation}
F_X (x) = \frac{1}{2}+ \mbox{sgn}(x-\mu)\frac{\kappa \big[1/\gamma,(\left| x-\mu \right| \eta(\sigma, \gamma))\gamma \big]}{2\Gamma(1/\gamma)},
\label{equ5}
\end{equation}
where $\eta \left( \sigma ,\gamma  \right)=\frac{1}{\sigma }\sqrt{\frac{\Gamma \left( 3/\gamma  \right)}{\Gamma \left( 1/\gamma  \right)}}$, $\kappa \left( \cdot  \right)$ is the lower incomplete gamma function, $\Gamma \left( \cdot  \right)$ is the gamma function, and $\text{sgn}\left( x \right)$ represents a symbol decision function, \ie, $\text{sgn}\left( x \right)=1$, if $x\ge 0$, and $\text{sgn}\left( x \right)=-1$ otherwise.

Now we introduce how to estimate three parameters of a GGD distribution from experimental results. For a PAM signal with order $M$, the transmitted signal is denoted by ${{x}_{i}}$, $\left( i=1,2,\ldots ,M \right)$, and the corresponding received signal
through a channel is denoted by ${{y}_{i}\left(j \right)}$, $\left( j=1,2,\ldots ,J \right)$, where $J$ is the total number of experimental results on each constellation point. Following \cite{Franklin2004Probability}, the sample mean ${{\mu }_{i}}$ and sample variance $\sigma _{i}^{2}$ of ${{y}_{i}}\left( j \right)$ are obtained as
\begin{equation}
{{\mu }_{i}}=\frac{1}{J}\underset{j=1}{\overset{J}{\mathop \sum }}\,{{y}_{i}}\left( j \right),
\label{equ6}
\end{equation}
\begin{equation}
\sigma _{i}^{2}=\frac{1}{J\text{-}1}\sum\limits_{j=1}^{J}{{{\big[{{y}_{i}}(j)-{{\mu }_{i}}\big]}^{2}}}.
\label{equ7}
\end{equation}

The estimation of the shape factor ${{\gamma }_{i}}$ is more difficult than the other two parameters. According to the results of \cite{Sharifi1995Estimation,4128961}, we obtain generalized Gaussian ratio function $r\left( {{\gamma }_{i}} \right)$ which is a function of ${{\gamma }_{i}}$ and is defined as
\begin{equation}
r\left( {{\gamma }_{i}} \right)=\frac{\sigma _{i}^{2}}{{{\left( \frac{1}{J}\underset{j=1}{\overset{J}{\mathop \sum }}\,\left| {{y}_{i}}(j)-{{\mu }_{i}}(j) \right| \right)}^{2}}}=\frac{\Gamma \left( 1/{{\gamma }_{i}} \right)\Gamma \left( 3/{{\gamma }_{i}} \right)}{{{\Gamma }^{2}}\left( 2/{{\gamma }_{i}} \right)}.
\label{equ8}
\end{equation}

By setting $r\left( {{\gamma }_{i}} \right)={{\rho }_{i}}$, a feasible solution of ${{\gamma }_{i}}$ can be found as
\begin{equation}
{{\gamma }_{i}}={{r}^{-1}}\left( {{\rho }_{i}} \right),
\label{equ9}
\end{equation}
where an exhausted search approach is employed for solving the (\ref{equ9}) to obtain an estimate of ${{\gamma }_{i}}$. Then, the value of ${{\gamma }_{i}}$ gradually increases from zero and the search process is completed until $r\left( {{\gamma }_{i}} \right)={{\rho }_{i}}$. Although the GGD has been also employed in \cite{Vill2006Multilevel} to model the channel of a $2$D barcode, there is a fundamental difference. It only considers the model of an SPS process rather than a DPS process in an illegal copying attack.

\section{Experiment Results}

\subsection{Experimental Results of Channel Modeling}
In our experiment, the parameters of our approach are given as follows: $k=440$ bits, $n=2040$ bits, $t=800$, ${{k}_{a}}=147$ bits, ${{n}_{a}}=255$ bits, ${{t}_{a}}=14$, $N=2$, ${{L}_{s}}=4080$ bits, ${{L}_{h}}=338\ $bits, and ${{L}_{t}}=4418$ bits.
Unless otherwise specified, our experiments follow these settings.
First, the printing material is chosen as the A$4$ paper with weight $120\text{g}/{{\text{m}}^{2}}$ from the Xerox.
Second, an original $2$D barcode with ${{L}_{t}}=47\times 47$ modules is printed on the chosen paper, where the printed size of each barcode is set as $3.2\times 3.2\text{ cm}^{2}$.
Last but not least, each combination is repeated $72$ times to obtain the average results.
The general experimental settings are summarized as follows:
\begin{itemize}
  \item Printing $1$: HP LaserJet P$1108$ printer in $1200$ DPI on paper with $120$ grams per square meter (gsm), and a rendering size of $3.2 \times 3.2$ cm$^2$;
  \item Printing $2$: FUJI P$355$D printer in $600$ DPI on paper with $120$ gsm, and a rendering size of $3.2 \times 3.2$ cm$^2$;
  \item Scanning $1$: BENQ K$810$ scanner in $1200$ DPI;
  \item Scanning $2$: EPSON V$330$ scanner in $600$ DPI;
  \item Camera Phone: HONOR V$20$ with $48$ MP resolution;
  \item Barcode Design: A multilevel barcode with $47 \times 47$ modules;
  \item Capture Angle: Within $10$ degrees between the barcode image plane and the camera sensor plane;
  \item Capture Distance: About $15$ cm in the in-focus case.
  \item Lighting: $300-350$ lux for the bright case and $100-150$ lux for the dim case.
\end{itemize}

Four metrics of bit error rates (BERs) can be calculated to characterize the reception performance against channel distortion. The first two metrics are used to measure the robustness of the authentication message. The first metric is the demodulated BER of the authentication message ${{\varepsilon }_{a,2}}$ which is calculated by comparing ${{\hat{s}}_{a,2}}$ and ${{s}_{a,2}}$, while the second metric is the decoded BER of the authentication message, ${{\varepsilon }_{a,1}}$, obtained by comparing ${{\hat{s}}_{a,1}}$ and ${{s}_{a,1}}$. Moreover, the remaining two metrics are used to represent the robustness of the source message, which also represents the covertness of the proposed LCAC code. The third metric is the demodulated BER of the source message, ${{\varepsilon }_{c,2}}$ computed by comparing ${{\hat{s}}_{c,2}}$ and ${{s}_{c,2}}$, while the fourth metric is the decoded BER of the source message, ${{\varepsilon }_{c,1}}$ by comparing ${{\hat{s}}_{c,1}}$ and ${{s}_{c,1}}$, and determined by the following formula: ${{\varepsilon }_{c,1}}={{\left| {{{\hat{\varepsilon }}}_{c,1}}-{{\varepsilon }_{c,1}} \right|}_{0}}/NK$.

\begin{table*}[!t]
\centering
\caption{Estimated parameters of a GGD approximation for illegally copied $2$D barcodes under 16 combinations.}
\footnotesize
\begin{minipage}[t]{0.22\linewidth}
\centering
(a) ${{\mathsf{P}}_{1}}-{{\mathsf{S}}_{1}}-{{\mathsf{P}}_{1}}-{{\mathsf{S}}_{1}}$
\begin{tabular}{|c|c|c|c|}
\hline
$x$ & $\mu \left( x \right)$ & ${{\sigma }^{2}}\left( x \right)$ & $\gamma \left( x \right)$ \\
\hline
40	&38.63	&138.53	&1.34\\
100	&119.87	&432.05	&1.77\\
160	&169.03	&333.62	&1.93\\
220	&215.83	&120.37	&1.76\\
\hline
\end{tabular}
\end{minipage}
\hspace{0.35cm}
\begin{minipage}[t]{0.22\linewidth}
\centering
(b) ${{\mathsf{P}}_{1}}-{{\mathsf{S}}_{1}}-{{\mathsf{P}}_{1}}-{{\mathsf{S}}_{2}}$
\begin{tabular}{|c|c|c|c|}
\hline
$x$ & $\mu \left( x \right)$ & ${{\sigma }^{2}}\left( x \right)$ & $\gamma \left( x \right)$ \\
\hline
40	&33.71	&115.09	&1.28\\
100	&106.77	&405.33	&1.73\\
160	&175.59	&242.49	&1.96\\
220	&211.03	&128.23	&1.91\\
\hline
\end{tabular}
\end{minipage}
\hspace{0.35cm}
\begin{minipage}[t]{0.22\linewidth}
\centering
(c) ${{\mathsf{P}}_{1}}-{{\mathsf{S}}_{1}}-{{\mathsf{P}}_{2}}-{{\mathsf{S}}_{1}}$
\begin{tabular}{|c|c|c|c|}
\hline
$x$ & $\mu \left( x \right)$ & ${{\sigma }^{2}}\left( x \right)$ & $\gamma \left( x \right)$ \\
\hline
40	&30.99	&156.18	&1.22\\
100	&110.31	&351.84	&1.75\\
160	&171.46	&217.05	&1.98\\
220	&210.87	&112.96	&1.99\\
\hline
\end{tabular}
\end{minipage}
\hspace{0.35cm}
\begin{minipage}[t]{0.22\linewidth}
\centering
(d) ${{\mathsf{P}}_{1}}-{{\mathsf{S}}_{1}}-{{\mathsf{P}}_{2}}-{{\mathsf{S}}_{2}}$
\begin{tabular}{|c|c|c|c|}
\hline
$x$ & $\mu \left( x \right)$ & ${{\sigma }^{2}}\left( x \right)$ & $\gamma \left( x \right)$ \\
\hline
40	&31.02	&124.53	&1.15\\
100	&107.25	&276.83	&1.67\\
160	&168.13	&252.68	&1.94\\
220	&218.78	&216.23	&1.96\\
\hline
\end{tabular}
\end{minipage}\\
\vspace{0.25cm}

\begin{minipage}[t]{0.22\linewidth}
\centering
(e) ${{\mathsf{P}}_{1}}-{{\mathsf{S}}_{2}}-{{\mathsf{P}}_{1}}-{{\mathsf{S}}_{1}}$
\begin{tabular}{|c|c|c|c|}
\hline
$x$ & $\mu \left( x \right)$ & ${{\sigma }^{2}}\left( x \right)$ & $\gamma \left( x \right)$ \\
\hline
40	&24.83	&34.01	&0.84\\
100	&100.54	&385.77	&1.83\\
160	&160.95	&267.73	&1.95\\
220	&202.44	&132.34	&1.81\\
\hline
\end{tabular}
\end{minipage}
\hspace{0.35cm}
\begin{minipage}[t]{0.22\linewidth}
\centering
(f) ${{\mathsf{P}}_{1}}-{{\mathsf{S}}_{2}}-{{\mathsf{P}}_{1}}-{{\mathsf{S}}_{2}}$
\begin{tabular}{|c|c|c|c|}
\hline
$x$ & $\mu \left( x \right)$ & ${{\sigma }^{2}}\left( x \right)$ & $\gamma \left( x \right)$ \\
\hline
40	&23.38	&24.83	&0.65\\
100	&97.87	&348.76	&1.73\\
160	&159.89	&273.04	&1.91\\
220	&207.91	&190.38	&1.91\\
\hline
\end{tabular}
\end{minipage}
\hspace{0.35cm}
\begin{minipage}[t]{0.22\linewidth}
\centering
(g) ${{\mathsf{P}}_{1}}-{{\mathsf{S}}_{2}}-{{\mathsf{P}}_{2}}-{{\mathsf{S}}_{1}}$
\begin{tabular}{|c|c|c|c|}
\hline
$x$ & $\mu \left( x \right)$ & ${{\sigma }^{2}}\left( x \right)$ & $\gamma \left( x \right)$ \\
\hline
40	&31.48	&29.65	&0.94\\
100	&105.64	&429.72	&2.35\\
160	&161.22	&221.04	&2.11\\
220	&203.75	&103.07	&1.72\\
\hline
\end{tabular}
\end{minipage}
\hspace{0.35cm}
\begin{minipage}[t]{0.22\linewidth}
\centering
(h) ${{\mathsf{P}}_{1}}-{{\mathsf{S}}_{2}}-{{\mathsf{P}}_{2}}-{{\mathsf{S}}_{2}}$
\begin{tabular}{|c|c|c|c|}
\hline
$x$ & $\mu \left( x \right)$ & ${{\sigma }^{2}}\left( x \right)$ & $\gamma \left( x \right)$ \\
\hline
40	&30.55	&23.02	&0.87\\
100	&100.79	&401.96	&2.26\\
160	&159.54	&272.79	&2.05\\
220	&214.87	&194.64	&1.71\\
\hline
\end{tabular}
\end{minipage}\\
\vspace{0.25cm}
\begin{minipage}[t]{0.22\linewidth}
\centering
(i) ${{\mathsf{P}}_{2}}-{{\mathsf{S}}_{1}}-{{\mathsf{P}}_{1}}-{{\mathsf{S}}_{1}}$
\begin{tabular}{|c|c|c|c|}
\hline
$x$ & $\mu \left( x \right)$ & ${{\sigma }^{2}}\left( x \right)$ & $\gamma \left( x \right)$ \\
\hline
40	&24.91	&85.63	&1.22\\
100	&114.92	&323.97	&2.16\\
160	&165.03	&226.41	&1.98\\
220	&198.55	&124.51	&1.99\\
\hline
\end{tabular}
\end{minipage}
\hspace{0.35cm}
\begin{minipage}[t]{0.22\linewidth}
\centering
(j) ${{\mathsf{P}}_{2}}-{{\mathsf{S}}_{1}}-{{\mathsf{P}}_{1}}-{{\mathsf{S}}_{2}}$
\begin{tabular}{|c|c|c|c|}
\hline
$x$ & $\mu \left( x \right)$ & ${{\sigma }^{2}}\left( x \right)$ & $\gamma \left( x \right)$ \\
\hline
40	&26.33	&72.71	&1.16\\
100	&114.82	&246.74	&2.09\\
160	&164.29	&198.53	&1.93\\
220	&201.43	&128.01	&2.01\\
\hline
\end{tabular}
\end{minipage}
\hspace{0.35cm}
\begin{minipage}[t]{0.22\linewidth}
\centering
(k) ${{\mathsf{P}}_{2}}-{{\mathsf{S}}_{1}}-{{\mathsf{P}}_{2}}-{{\mathsf{S}}_{1}}$
\begin{tabular}{|c|c|c|c|}
\hline
$x$ & $\mu \left( x \right)$ & ${{\sigma }^{2}}\left( x \right)$ & $\gamma \left( x \right)$ \\
\hline
40	&29.21	&110.79	&1.38\\
100	&111.27	&238.01	&2.07\\
160	&166.02	&158.36	&1.99\\
220	&188.91	&76.36	&1.86\\
\hline
\end{tabular}
\end{minipage}
\hspace{0.35cm}
\begin{minipage}[t]{0.22\linewidth}
\centering
(l) ${{\mathsf{P}}_{2}}-{{\mathsf{S}}_{1}}-{{\mathsf{P}}_{2}}-{{\mathsf{S}}_{2}}$
\begin{tabular}{|c|c|c|c|}
\hline
$x$ & $\mu \left( x \right)$ & ${{\sigma }^{2}}\left( x \right)$ & $\gamma \left( x \right)$ \\
\hline
40	&33.69	&79.32	&1.44\\
100	&103.29	&198.15	&2.01\\
160	&165.97	&294.82	&1.84\\
220	&202.22	&220.46	&1.88\\
\hline
\end{tabular}
\end{minipage}\\
\vspace{0.25cm}
\begin{minipage}[t]{0.22\linewidth}
\centering
(m) ${{\mathsf{P}}_{2}}-{{\mathsf{S}}_{2}}-{{\mathsf{P}}_{1}}-{{\mathsf{S}}_{1}}$
\begin{tabular}{|c|c|c|c|}
\hline
$x$ & $\mu \left( x \right)$ & ${{\sigma }^{2}}\left( x \right)$ & $\gamma \left( x \right)$ \\
\hline
40	&25.41	&31.56	&0.96\\
100	&111.59	&217.18	&1.88\\
160	&160.26	&172.17	&1.92\\
220	&198.09	&124.09	&2.02\\
\hline
\end{tabular}
\end{minipage}
\hspace{0.35cm}
\begin{minipage}[t]{0.22\linewidth}
\centering
(n) ${{\mathsf{P}}_{2}}-{{\mathsf{S}}_{2}}-{{\mathsf{P}}_{1}}-{{\mathsf{S}}_{2}}$
\begin{tabular}{|c|c|c|c|}
\hline
$x$ & $\mu \left( x \right)$ & ${{\sigma }^{2}}\left( x \right)$ & $\gamma \left( x \right)$ \\
\hline
40	&25.14	&24.09	&0.82\\
100	&105.02	&177.93	&1.82\\
160	&156.64	&212.76	&1.89\\
220	&209.84	&276.31	&1.91\\
\hline
\end{tabular}
\end{minipage}
\hspace{0.35cm}
\begin{minipage}[t]{0.22\linewidth}
\centering
(o) ${{\mathsf{P}}_{2}}-{{\mathsf{S}}_{2}}-{{\mathsf{P}}_{2}}-{{\mathsf{S}}_{1}}$
\begin{tabular}{|c|c|c|c|}
\hline
$x$ & $\mu \left( x \right)$ & ${{\sigma }^{2}}\left( x \right)$ & $\gamma \left( x \right)$ \\
\hline
40	&30.64	&21.97	&0.85\\
100	&103.56	&215.04	&2.24\\
160	&152.81	&183.35	&2.03\\
220	&187.54	&139.71	&2.17\\
\hline
\end{tabular}
\end{minipage}
\hspace{0.35cm}
\begin{minipage}[t]{0.22\linewidth}
\centering
(p) ${{\mathsf{P}}_{2}}-{{\mathsf{S}}_{2}}-{{\mathsf{P}}_{2}}-{{\mathsf{S}}_{2}}$
\begin{tabular}{|c|c|c|c|}
\hline
$x$ & $\mu \left( x \right)$ & ${{\sigma }^{2}}\left( x \right)$ & $\gamma \left( x \right)$ \\
\hline
40	&22.11	&22.59	&0.64\\
100	&108.63	&260.99	&2.18\\
160	&163.53	&208.76	&1.99\\
220	&196.91	&139.13	&2.19\\
\hline
\end{tabular}
\end{minipage}
\label{Table2}
\end{table*}

\begin{figure*}[!t]
\centering
\includegraphics[width=.99\linewidth]{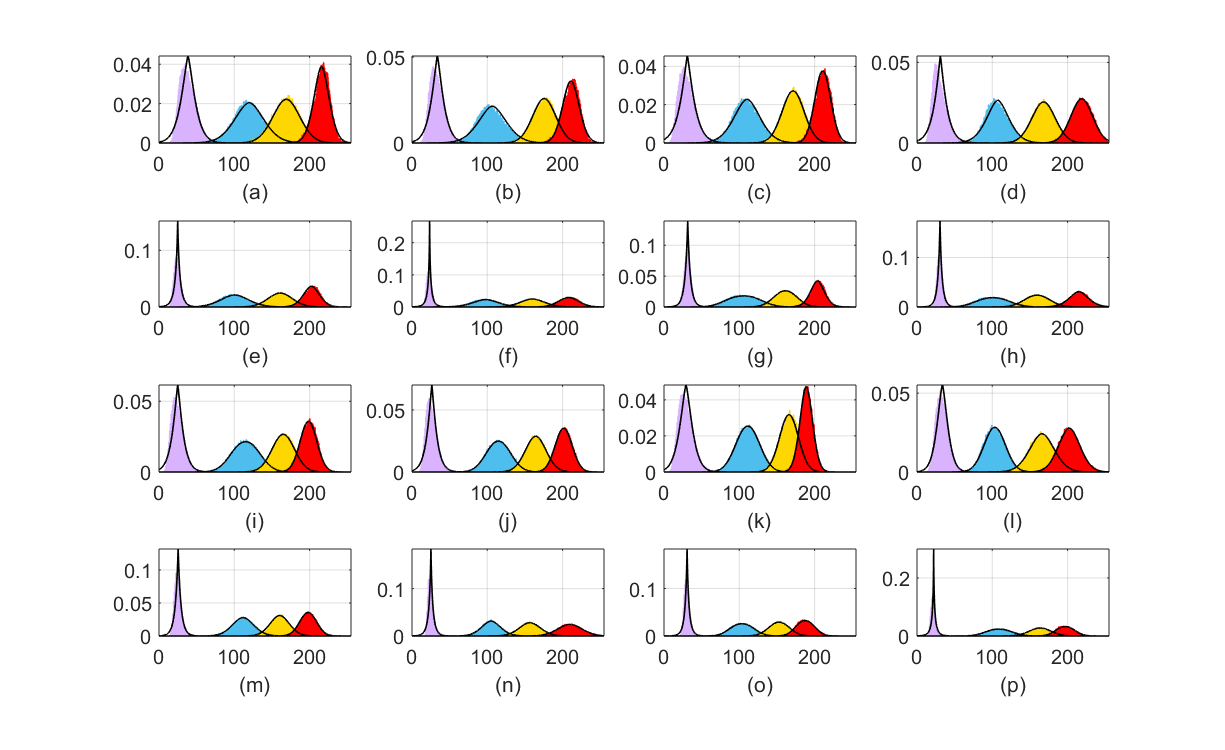}
\caption{Histograms of the received intensities in an illegally copied $2$D barcode and the approximation curves of GGD models under $16$ combinations: (a) ${{\mathsf{P}}_{1}}-{{\mathsf{S}}_{1}}-{{\mathsf{P}}_{1}}-{{\mathsf{S}}_{1}}$; (b) ${{\mathsf{P}}_{1}}-{{\mathsf{S}}_{1}}-{{\mathsf{P}}_{1}}-{{\mathsf{S}}_{2}}$; (c) ${{\mathsf{P}}_{1}}-{{\mathsf{S}}_{1}}-{{\mathsf{P}}_{2}}-{{\mathsf{S}}_{1}}$; (d) ${{\mathsf{P}}_{1}}-{{\mathsf{S}}_{1}}-{{\mathsf{P}}_{2}}-{{\mathsf{S}}_{2}}$; (e) ${{\mathsf{P}}_{1}}-{{\mathsf{S}}_{2}}-{{\mathsf{P}}_{1}}-{{\mathsf{S}}_{1}}$; (f) ${{\mathsf{P}}_{1}}-{{\mathsf{S}}_{2}}-{{\mathsf{P}}_{1}}-{{\mathsf{S}}_{2}}$; (g) ${{\mathsf{P}}_{1}}-{{\mathsf{S}}_{2}}-{{\mathsf{P}}_{2}}-{{\mathsf{S}}_{1}}$; (h) ${{\mathsf{P}}_{1}}-{{\mathsf{S}}_{2}}-{{\mathsf{P}}_{2}}-{{\mathsf{S}}_{2}}$; (i) ${{\mathsf{P}}_{2}}-{{\mathsf{S}}_{1}}-{{\mathsf{P}}_{1}}-{{\mathsf{S}}_{1}}$; (j) ${{\mathsf{P}}_{2}}-{{\mathsf{S}}_{1}}-{{\mathsf{P}}_{1}}-{{\mathsf{S}}_{2}}$; (k) ${{\mathsf{P}}_{2}}-{{\mathsf{S}}_{1}}-{{\mathsf{P}}_{2}}-{{\mathsf{S}}_{1}}$; (l) ${{\mathsf{P}}_{2}}-{{\mathsf{S}}_{1}}-{{\mathsf{P}}_{2}}-{{\mathsf{S}}_{2}}$; (m) ${{\mathsf{P}}_{2}}-{{\mathsf{S}}_{2}}-{{\mathsf{P}}_{1}}-{{\mathsf{S}}_{1}}$; (n) ${{\mathsf{P}}_{2}}-{{\mathsf{S}}_{2}}-{{\mathsf{P}}_{1}}-{{\mathsf{S}}_{2}}$; (o) ${{\mathsf{P}}_{2}}-{{\mathsf{S}}_{2}}-{{\mathsf{P}}_{2}}-{{\mathsf{S}}_{1}}$; (p) ${{\mathsf{P}}_{2}}-{{\mathsf{S}}_{2}}-{{\mathsf{P}}_{2}}-{{\mathsf{S}}_{2}}$.}
\label{Figure8}
\vspace{-0.5cm}
\end{figure*}
According to (\ref{equ6}), (\ref{equ7}) and (\ref{equ9}), the estimation results of three parameters of illegally copied $2$D barcode under $16$ combinations are given in Tab. \ref{Table2}. We accumulated the $72$ samples obtained from each combination, and the blocks corresponding to the four gray values were superimposed respectively to calculate the frequency of the actual gray values. Then the corresponding histogram of the received signal constellations in an illegally copied $2$D barcode and GGD approximation are shown in Fig. \ref{Figure8}. We can see that the experimental results match well with a GGD approximation for all constellations except ${{x}_{1}}=40$. It is conjectured that since the points of ${{x}_{1}}=40$ have the lowest intensity level as compared with the other constellations, which is more sensitive to the distortions in an illegal copying process \cite{zhang2019accurate}.

Moreover, from Fig. \ref{Figure8}, we can also obtain the following observations. First, by comparing the results of the subfigures in the first two rows of Fig. \ref{Figure8} with those in the last two rows, we find that the sample mean of ${{x}_{4}}=220$ in the former group is larger than those in the latter group, which is due to the different printers in the first printing process. Second, by comparing the results of the subfigures in the first and the third rows of Fig. \ref{Figure8} with those in the second and fourth rows, we find that the peak value of histogram at ${{x}_{1}}=40$ in the former group is smaller than those in the latter group, \ie, about $0.05$ and $0.15$, which is determined by the chosen scanner in the first scanning process; Finally, by comparing the results of the subfigures in the first and third columns of Fig. \ref{Figure8} with those in the second and fourth columns, we find that the peak value of histogram at ${{x}_{1}}=40$ in the former group is smaller than those in the latter group, which is determined by the chosen scanner in the second scanning process.

\subsection{Advanced Illegal Copying Strategy}
The experimental results given in the previous subsection are based on a simple illegal copying strategy, \ie, direct-copying (DC) attack. If an attacker can capture multiple printed samples of the same legal $2$D barcode, he or she can utilize all samples to improve the probability of a successful attack. For example, the attacker first generates a synthesized $2$D barcode with better quality by averaging the intensities over all received barcode samples and then illegally prints it. This strategy is termed as a synthesized-copying (SC) attack.

For convenience, the device combination of ${{\mathsf{P}}_{1}}\mathsf{-}{{\mathsf{S}}_{1}}\mathsf{-}{{\mathsf{P}}_{1}}\mathsf{-}{{\mathsf{S}}_{1}}$ is chosen to emulate an SC attack. We still use a GGD to model an SC attack, \ie, $\left( \mathcal{G}\mathcal{G}\mathcal{D}\left( {{\mu }_{i}}\left( {{n}_{s}} \right),\sigma _{i}^{2}\left( {{n}_{s}} \right),{{\gamma }_{i}}\left( {{n}_{s}} \right) \right) \right)$, where ${{n}_{s}}$ is the number of synthesized samples. Similar to the equations from (\ref{equ6}) to (\ref{equ9}), three parameters are obtained as
\begin{equation}
{{\mu }_{i}}\left( {{n}_{s}} \right)=\frac{1}{J}\underset{j=1}{\overset{J}{\mathop \sum }}\,\left( \frac{1}{{{n}_{s}}}\sum\limits_{s=1}^{{{n}_{s}}}{{{y}_{i}}\left( s,j \right)} \right),
\label{equ10}
\end{equation}
\begin{equation}
\sigma _{i}^{2}\left( {{n}_{s}} \right)=\frac{1}{J-1}{{\sum\limits_{j=1}^{J}{\left( \Big( \frac{1}{{{n}_{s}}}\sum\limits_{s=1}^{{{n}_{s}}}{{{y}_{i}}(s,j)} \Big)-{{\mu }_{i}}\left( {{n}_{s}} \right) \right)}}^{2}},
\label{equ11}
\end{equation}
\begin{equation}
{{\gamma }_{i}}\left( {{n}_{s}} \right)={{r}^{-1}}\left( {{\rho }_{i}}\left( {{n}_{s}} \right) \right),
\label{equ12}
\end{equation}
where
\begin{equation}
{{\rho }_{i}}\left( {{n}_{s}} \right)=\frac{\sigma _{i}^{2}\left( {{n}_{s}} \right)}{{{\left( \frac{1}{J-1}\underset{j=1}{\overset{J}{\mathop \sum }}\,\left| \left( \frac{1}{{{n}_{s}}}\sum\limits_{s=1}^{{{n}_{s}}}{{{y}_{i}}(s,j)} \right)-{{\mu }_{i}}({{n}_{s}}) \right| \right)}^{2}}}.
\label{equ13}
\end{equation}
Note that if ${{n}_{s}}=1$, an SC attack reduces to a DC attack.
\subsubsection{Experimental Results at an Attack's Receiver}
\begin{figure}[!t]
\centering
\includegraphics[height=6cm]{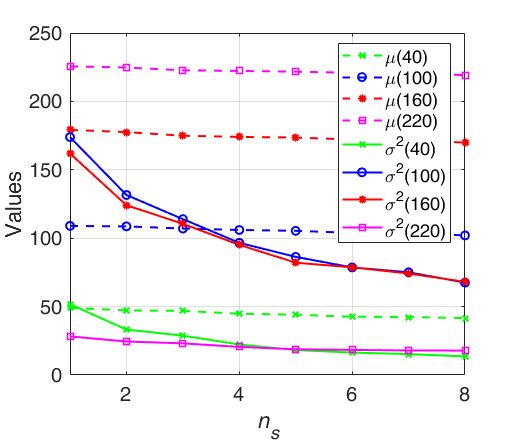}
\caption{Estimated parameters of an SC attack for an attacker's receiver.}
\label{Figure9}
\end{figure}
The estimated parameters of an SC attack with (\ref{equ10}), (\ref{equ11}) and (\ref{equ12}) are presented in Fig. \ref{Figure9}. We can see that as ${{n}_{s}}$ increases, the sample mean gradually closes the corresponding ideal constellation point, and the sample variance decreases as expected. However, the decrease rates of the variance become slower gradually. For example, by comparing the case of ${{n}_{s}}=1$ for ${{x}_{4}}=220$ with that of ${{n}_{s}}=2$, the variance is decreased from 28.20 to 24.48 and the decreasing ratio is $\left( 28.20 - 24.48 \right)/28.20=13.19\%$. In contrast, by comparing the case of ${{n}_{s}}=7$ for ${{x}_{4}}=220$ with that of ${{n}_{s}}=8$, the variance is only decreased from $17.94$ to $17.84$ and the decreasing ratio is $\left( 17.94 - 17.84 \right)/17.94=0.56\%$. Thus, we can draw an important conclusion of an SC attack: although a synthesized operation can improve attack accuracy, this improvement gradually approaches a bottleneck.

\subsubsection{Experimental Results at a Legal Receiver}
\begin{figure}[!t]
\centering
\includegraphics[height=6cm]{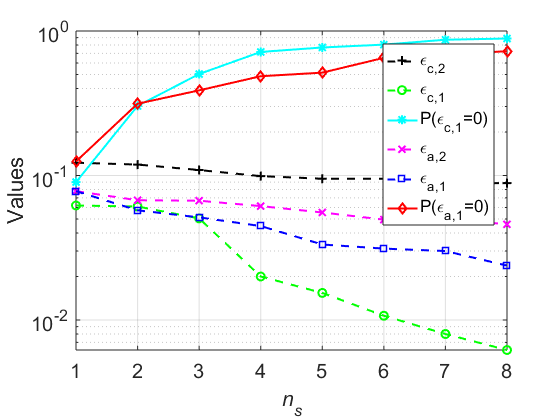}
\caption{BERs analysis of an SC attack for a legal receiver (Scanner).}
\label{Figure10}
\end{figure}

\begin{figure}[!t]
\centering
\includegraphics[height=6cm]{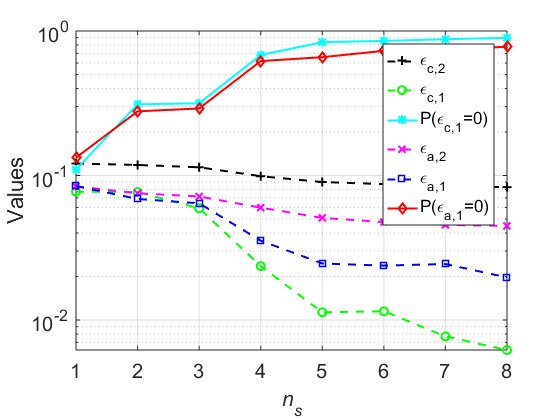}
\caption{BERs analysis of an SC attack for a legal receiver (Mobile Phone).}
\label{Figure11}
\end{figure}

Through a synthesized operation, an attacker generates $2$D barcode and prints it on a paper. Then, the legal receiver can scan it and perform authentication to detect an illegal copying $2$D barcode.
Now, we present the BERs analysis of an SC attack for a legal receiver, where both scanner and mobile phone are considered as the capturing device of a legal receiver, as shown in Fig. \ref{Figure10} and Fig. \ref{Figure11}, respectively.
From Fig. \ref{Figure10} and Fig. \ref{Figure11}, we can see that as ${{n}_{s}}$ increases, the values of all BERs are decreased as expected.
At the same time, the probability that the BERs is equal to $0$ also increases gradually with the increase of ${{n}_{s}}$.
For example, as shown in Fig. \ref{Figure10}, when ${{n}_{s}}=1$, $\mathbb{P}\left( {{\varepsilon }_{c,1}}=0 \right)=0.0909$ and $\mathbb{P}\left( {{\varepsilon }_{a,1}}=0 \right)=0.1250$; when ${{n}_{s}}=8$, $\mathbb{P}\left( {{\varepsilon }_{c,1}}=0 \right)=0.8871$ and $\mathbb{P}\left( {{\varepsilon }_{a,1}}=0 \right)=0.7258$.
Here, $\mathbb{P}\left( \cdot  \right)$ denotes a probability measure.
Moreover, by compare the results of Fig. \ref{Figure11} with those of Fig. \ref{Figure10}, we find that the results under a mobile phone is better than those under a scanner.
For example, as shown in Fig. \ref{Figure11}, when ${{n}_{s}}=1$, $\mathbb{P}\left( {{\varepsilon }_{c,1}}=0 \right)=0.1102$ and $\mathbb{P}\left( {{\varepsilon }_{a,1}}=0 \right)=0.1338$; when ${{n}_{s}}=8$, $\mathbb{P}\left( {{\varepsilon }_{c,1}}=0 \right)=0.8978$ and $\mathbb{P}\left( {{\varepsilon }_{a,1}}=0 \right)=0.7790$.
The advantage under a mobile phone is because the selected mobile phone has better capturing resolution than that under the selected scanner, which is further verified in the next section by comparing the bias from the standard constellation point to the recovered constellation point.
Note that, since the channel distortions in a DPS process are more severe than those of an SPS process, the legal receiver cannot always decode both the source and authentication messages without errors.

\subsubsection{Verification Decision in the Proposed LCAC code}
Note that the main objective of this paper is to find an effective and low-cost approach to authenticate an illegal copying $2$D barcode.
Based on the experimental results of both Fig. \ref{Figure10} and Fig. \ref{Figure11}, we consider two authentication approaches.
In the first approach, the authentication decision is made by checking whether the source message can be ideally decoded, \ie, ${{\varepsilon }_{c,1}}=0$.
In the second approach, the authentication decision is made by checking whether the authentication message can be ideally decoded, \ie, ${{\varepsilon }_{a,1}}=0$.
Thus, the results of both $\mathbb{P}\left( {{\varepsilon }_{c,1}}=0 \right)$ and $\mathbb{P}\left( {{\varepsilon }_{a,1}}=0 \right)$ are also provided in both Fig. \ref{Figure10} and Fig. \ref{Figure11}.

Although the first authentication approach is simple even no embedding of authentication message, it has two obvious drawbacks. First, as ${{n}_{s}}$ increases, the value of $\mathbb{P}\left( {{\varepsilon }_{c,1}}=0 \right)$ is obviously increased, which lowers the efficiency of the first approach. Second, the parameters of a source message are predetermined according to certain considered of a $2$D barcode, and it cannot be arbitrarily changed for improving the authentication accuracy. In contrary, the second authentication approach is a better option, although the value of $\mathbb{P}\left( {{\varepsilon }_{a,1}}=0 \right)$ is also increased as ${{n}_{s}}$ increases. This is because the parameters of an authentication message are freely adjustable to improve the authentication accuracy, which overcomes the second drawback of the first authentication approach and is the most attractive feature of the proposed LCAC code. Therefore, we use the second authentication approach as the authentication decision block of the receiver in the proposed LCAC code, as shown in Fig. \ref{Figure6}, which is specifically described as the following definition.

\textbf{Definition 1}: The authentication decision block of the receiver in the proposed LCAC code identifies the captured $2$D barcode as an illegal copying one, if ${{\varepsilon }_{a,1}}>\delta $, where $0\le \delta <1$ is a threshold of the authentication decision.

Note that a smaller value of $\delta$ corresponds to higher security, \eg, $\delta =0$ indicates that the captured $2$D barcode is identified as an illegal copying one if there is any decoded error. However, a false rejection decision may accidentally occur, since there is also a decoded error (${{\varepsilon }_{a,1}}\ne 0$) for a legal $2$D barcode under some unideal situations, \eg, the resolution of a capturing device is not sufficiently high or the lighting is poor. Thus, the value of $\delta$ should not be set too small to avoid false decision. On the contrary, the value of $\delta$ also should not be set too large, otherwise, it will increase the success possibility of illegally-copying attacks. In the next section, the criterion for selecting $\delta$ will be discussed based on the chosen device and the parameters of the proposed LCAC code can then be optimized.

\section{Optimization of Embedding Parameters}
In this section, we optimize the parameters of the proposed LCAC $2$D barcode in order to increase the cost of copying attack and achieve a better anti-copying effect, which consists of four steps.
First, the experimental results of modeling for an SC attack is presented.
Second, based on the modeling results, a prediction function is established.
Third, based on the prediction function, the parameters of the proposed LCAC code will be optimized.
Finally, the selection of $\delta$ is discussed and experimental results are given.

\subsection{Experimental Results of Modeling for an SC Attack}

\begin{table*}[!t]
\centering
\caption{Estimated parameters of a GGD approximation for an SC attack under 8 cases (Mobile Phone).}
\footnotesize
\begin{minipage}[t]{0.22\linewidth}
\centering
(a) ${{n}_{s}}=1$
\begin{tabular}{|c|c|c|c|}
\hline
$x$ & $\mu \left( x \right)$ & ${{\sigma }^{2}}\left( x \right)$ & $\gamma \left( x \right)$ \\
\hline
40	&41.06	&119.24	&1.39\\
100	&117.61	&474.53	&1.91\\
160	&167.05	&378.64	&2.18\\
220	&214.92	&129.37	&1.76\\
\hline
\end{tabular}
\end{minipage}
\hspace{0.35cm}
\begin{minipage}[t]{0.22\linewidth}
\centering
(b) ${{n}_{s}}=2$
\begin{tabular}{|c|c|c|c|}
\hline
$x$ & $\mu \left( x \right)$ & ${{\sigma }^{2}}\left( x \right)$ & $\gamma \left( x \right)$ \\
\hline
40	&40.45	&94.92	&1.39\\
100	&116.55	&470.69	&1.72\\
160	&171.53	&348.23	&2.04\\
220	&218.60	&113.73	&1.79\\
\hline
\end{tabular}
\end{minipage}
\hspace{0.35cm}
\begin{minipage}[t]{0.22\linewidth}
\centering
(c) ${{n}_{s}}=3$
\begin{tabular}{|c|c|c|c|}
\hline
$x$ & $\mu \left( x \right)$ & ${{\sigma }^{2}}\left( x \right)$ & $\gamma \left( x \right)$ \\
\hline
40	&40.66	&92.70	&1.33\\
100	&116.17	&431.69	&1.79\\
160	&165.71	&345.62	&1.98\\
220	&212.48	&126.68	&1.81\\
\hline
\end{tabular}
\end{minipage}
\hspace{0.35cm}
\begin{minipage}[t]{0.22\linewidth}
\centering
(d) ${{n}_{s}}=4$
\begin{tabular}{|c|c|c|c|}
\hline
$x$ & $\mu \left( x \right)$ & ${{\sigma }^{2}}\left( x \right)$ & $\gamma \left( x \right)$ \\
\hline
40	&38.77	&78.34	&1.33\\
100	&112.63	&418.92	&1.87\\
160	&163.50	&325.93	&1.78\\
220	&212.21	&121.21	&1.72\\
\hline
\end{tabular}
\end{minipage}\\
\vspace{0.25cm}
\begin{minipage}[t]{0.22\linewidth}
\centering
(e) ${{n}_{s}}=5$
\begin{tabular}{|c|c|c|c|}
\hline
$x$ & $\mu \left( x \right)$ & ${{\sigma }^{2}}\left( x \right)$ & $\gamma \left( x \right)$ \\
\hline
40	&39.76	&71.82	&1.44\\
100	&111.42	&414.32	&1.65\\
160	&163.70	&315.48	&1.92\\
220	&212.41	&106.46	&1.81\\
\hline
\end{tabular}
\end{minipage}
\hspace{0.35cm}
\begin{minipage}[t]{0.22\linewidth}
\centering
(f) ${{n}_{s}}=6$
\begin{tabular}{|c|c|c|c|}
\hline
$x$ & $\mu \left( x \right)$ & ${{\sigma }^{2}}\left( x \right)$ & $\gamma \left( x \right)$ \\
\hline
40	&39.94	&73.17	&1.41\\
100	&111.09	&412.73	&1.66\\
160	&163.81	&307.67	&1.94\\
220	&212.76	&97.87	&1.71\\
\hline
\end{tabular}
\end{minipage}
\hspace{0.35cm}
\begin{minipage}[t]{0.22\linewidth}
\centering
(g) ${{n}_{s}}=7$
\begin{tabular}{|c|c|c|c|}
\hline
$x$ & $\mu \left( x \right)$ & ${{\sigma }^{2}}\left( x \right)$ & $\gamma \left( x \right)$ \\
\hline
40	&39.14	&79.49 &1.42\\
100	&111.14	&400.48	&1.64\\
160	&164.69	&292.47	&1.95\\
220	&214.03	&93.28	&1.72\\
\hline
\end{tabular}
\end{minipage}
\hspace{0.35cm}
\begin{minipage}[t]{0.22\linewidth}
\centering
(h) ${{n}_{s}}=8$
\begin{tabular}{|c|c|c|c|}
\hline
$x$ & $\mu \left( x \right)$ & ${{\sigma }^{2}}\left( x \right)$ & $\gamma \left( x \right)$ \\
\hline
40	&38.25	&71.40	&1.26\\
100	&110.30	&393.58	&1.61\\
160	&164.88 &282.19	&1.83\\
220	&213.33	&87.11	&1.63\\
\hline
\end{tabular}
\end{minipage}
\label{Table4}
\end{table*}


\begin{figure*}[!t]
\centering
\includegraphics[width=.99\linewidth]{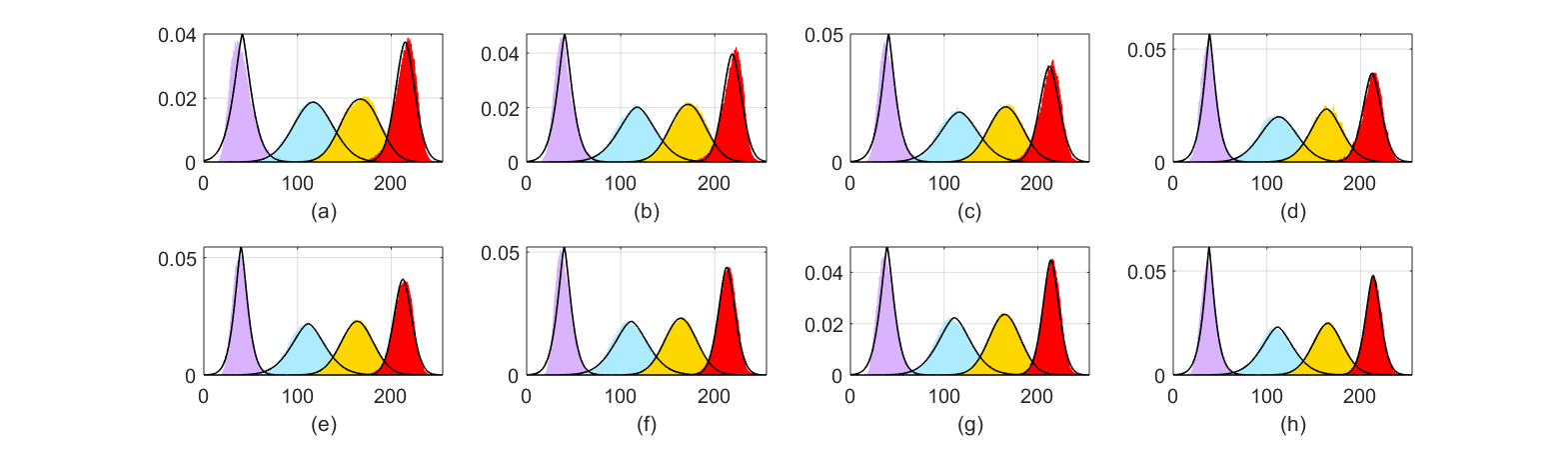}
\caption{Histograms of the received intensities in an illegally copied $2$D barcode and the approximation curves of GGD models for an SC attack under $8$ cases: (a) ${{n}_{s}}=1$; (b) ${{n}_{s}}=2$; (c) ${{n}_{s}}=3$; (d) ${{n}_{s}}=4$; (e) ${{n}_{s}}=5$; (f) ${{n}_{s}}=6$; (g) ${{n}_{s}}=7$; (h) ${{n}_{s}}=8$ (Mobile Phone).}
\label{Figure13}
\vspace{-0.5cm}
\end{figure*}

Although the modeling results of an SC attack given in the previous subsection work well, it is mathematically intractable to obtain an exact theoretical expression for the sum of multiple GGD RVs. Thus, we propose a simple method to simplify the process of establishing the prediction function. First, following \cite{Soury2015New}, we assume the sum of multiple GGD RVs can be approximated by a new GGD RV, \ie, $\mathcal{G}\mathcal{G}\mathcal{D}( {{{\hat{\mu }}}_{f}},\hat{\sigma }_{f}^{2},{{{\hat{\gamma }}}_{f}} )$. Then the three parameters of the new GGD RV can be estimated as follows. It should be noted that we have fitted all three parameters of the curve, which is a process of correction. According to the actual fluctuations of each parameter, we used a variety of common curve fitting functions (Exponential, Gaussian, Linear Fitting, Polynomial, Power with one term or two terms and so on) for correction, and finally chose one of them with the smallest error.

According to (\ref{equ10}), (\ref{equ11}) and (\ref{equ12}), the estimates of three parameters of a GGD approximation for an SC attack under $8$ cases under a mobile phone are given in Tab. \ref{Table4}.
Then the corresponding normalized histogram of the received signal points in an illegally copied $2$D barcode and GGD approximation for an SC attack are given in Fig. \ref{Figure13}.
Similar to Fig. \ref{Figure8}, Fig. \ref{Figure13} shows that the experimental results match well with a GGD approximation for all constellation points except ${{x}_{1}}=40$.
As ${{n}_{s}}$ increases, we obtain two conclusions.
First, the sample mean of all received signal points gradually close the ideal constellation point.
Second, the sample variance of all received signal points gradually decreases except that ${{x}_{4}}=220$, which indicates the complexity of a DPS process and the model is more accurate in the mid-range region of the histogram.

\subsection{Prediction Function}
In the previous section, through experimental results, we find that as ${{n}_{s}}$ increases, the success possibility of an SC attack is increased. It is desirable to optimize the parameters of the proposed LCAC code (${{k}_{a}}$, ${{n}_{a}}$ and $\delta$) to lower the success possibility of an SC attack. In other words, the attacker has to increase the value of ${{n}_{s}}$ to satisfy the condition of authentication decision, \ie, ${{\varepsilon }_{a,1}}>\delta $. The attacker is required to obtain more numbers of legal $2$D barcodes from the manufacturer, which significantly increases the cost of illegally-copying attacks. If the manufacturer can predict the value of ${{n}_{s}}$, the used times of a $2$D barcode can be determined to make a tradeoff between the costs of production and illegal copying. Specifically, a larger ${{n}_{s}}$ corresponds to a lower production cost but increase the cost of illegally-copying attacks and vice versa. An extreme case is that if the $2$D barcode generated by the merchant is unique, then his production cost is very high, and the return is that the attacker cannot generate an illegal $2$D barcode through synthetic attacks. So we need to make a trade-off to estimate a reasonable ${{n}_{s}}$.

It is tedious and difficult to predict the value of ${{n}_{s}}$ by modeling an SC attack based on various experiments and all numbers of synthesized samples. It is even impossible under certain situation, \eg, the device used by an attacker is unknown to the manufacturer. Thus, this paper considers a feasible solution to predict the value of ${{n}_{s}}$. Specifically, we first establish a prediction function based on the modeling results of an SC attack with a finite number of synthesized samples. Then, based on the prediction function, the manufacturer can effectively estimate the value of ${{n}_{s}}$.

First, we use a power fitting function with two variables to estimate ${{\hat{\mu }}_{f}}$ as
\begin{equation}
{{\hat{\mu }}_{f}}={{a}_{\mu }}{{f}^{{{b}_{\mu }}}},
\label{equ14}
\end{equation}
where $f$ represents the estimated number of synthesized samples, two variables (${{a}_{\mu }}$ and ${{b}_{\mu }}$) are obtained according to ${{\mu }_{i~}}\left( {{n}_{s}} \right)={{a}_{\mu }}n_{s}^{{{b}_{\mu }}}$.

Second, we use a power fitting function with three variables to estimate $\hat{\sigma }_{f}^{2}$ as
\begin{equation}
\hat{\sigma }_{f}^{2}={{a}_{\sigma }}{{f}^{{{b}_{\sigma }}}}+{{c}_{\sigma }},
\label{equ15}
\end{equation}
where three variables (${{a}_{\sigma }}$, ${{b}_{\sigma }}$, and ${{c}_{\sigma }}$) are obtained according to $\sigma _{i}^{2}\left( {{n}_{s}} \right)={{a}_{\sigma }}n_{s}^{{{b}_{\sigma }}}+{{c}_{\sigma }}$.

Third, since from Tab. \ref{Table4}., we can see that the value of the shape factor fluctuates around a certain constant, we directly use an average fitting function to estimate ${{\hat{\gamma }}_{f}}$ as
\begin{equation}
{{\hat{\gamma }}_{f}}=\frac{1}{{{n}_{s}}}\sum\limits_{s=1}^{{{n}_{s}}}{\gamma \left( s \right)}.
\label{equ16}
\end{equation}

\begin{figure*}[!t]
\centering
\includegraphics[width=\linewidth]{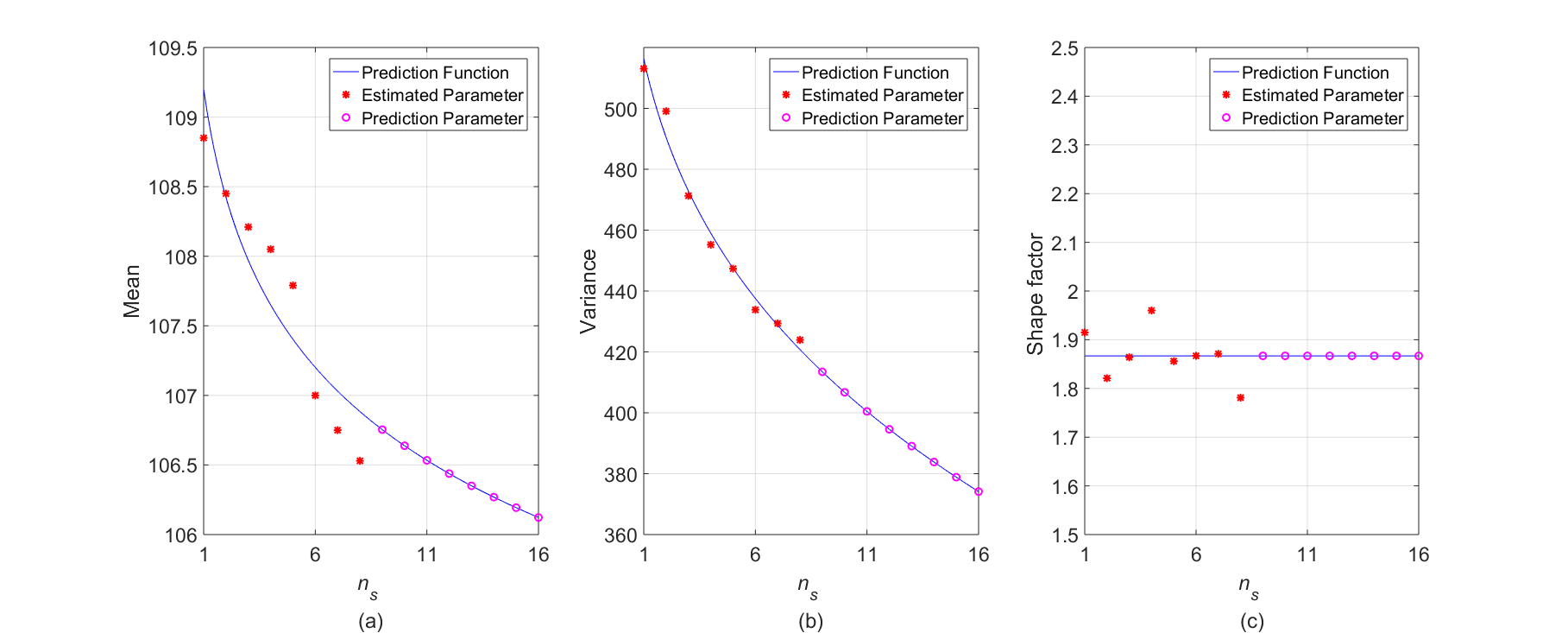}
\caption{Comparisons between the three estimated parameters and predicted parameters for ${{x}_{2}}=100$: (a) mean; (b) variance; (c) shape factor (Mobile Phone).}
\label{Figure15}
\vspace{-0.5cm}
\end{figure*}
%
Based on the results of Tab. \ref{Table4}, prediction functions under a mobile phone are devised.
For the mobile phone $\mathsf{M}$, the estimated parameters and predicted parameters for ${{x}_{2}}=100$ are illustrated in Fig. \ref{Figure15}, where ${{a}_{\mu }}=118.6$, ${{b}_{\mu }}=-0.0342$, ${{a}_{\sigma }}=-138.5$, ${{b}_{\sigma }}=0.232$, and ${{c}_{\sigma }}=617.7$.
From Fig. \ref{Figure15}, we can see that the estimated parameters oscillate around the predicted parameters, which verifies the efficacy of the prediction function.

\subsection{Parameter Optimization}
Based on a prediction function $\mathcal{GGD} ( {{{\hat{\mu }}}_{f}},\hat{\sigma }_{f}^{2},{{{\hat{\gamma }}}_{f}} )$, we can predict ${{\hat{\varepsilon }}_{a,2}}$ and even ${{\hat{\varepsilon }}_{a,1}}$. Due to the space limitation, we briefly introduce how to obtain ${{\hat{\varepsilon }}_{a,2}}$. First, the BER of certain constellation point ${{x}_{i}}$ can be calculated as
\begin{equation}
{{\hat{\varepsilon }}_{i}}=\sum\limits_{j=0,j+1\ne i}^{M-1}{{{\alpha }_{j}}\Big( {{F}_{i}}\left( {{\theta }_{j+1}} \right) - {{F}_{i}}\left( {{\theta }_{j}} \right) \Big)},
\label{equ17}
\end{equation}
where ${{\theta }_{j}}$   represents decision threshold, \eg, the case of $M=4$, ${{\theta }_{0}}=0$, ${{\theta }_{1}}=70$, ${{\theta }_{2}}=130$, ${{\theta }_{3}}=190$, and ${{\theta }_{4}}=255$, ${{\alpha }_{j}}$ is a correction factor, \eg, the case of $M=4$, ${{\alpha }_{j}}=\frac{1}{2}$ for $\left| j+1-i \right|\le 2$; otherwise ${{\alpha }_{j}}=1$, ${{F}_{i}}\left( \theta  \right)$ represents a CDF based on the predicted parameters, expressed as
\begin{equation}
F_X(x_i) = \frac{1}{2} + \mbox{sgn}(x_i - \hat{\mu}_f) \frac{\kappa \big[ 1/\hat{\gamma}_f, (\left| x_i - \hat{\mu}_f \right| \eta(\hat{\sigma}_f, \hat{\gamma}_f)) \hat{\gamma}_f \big]}{2\Gamma(1/\hat{\gamma}_f)}.
\label{equ18}
\end{equation}
Then the theoretical expressions of ${{\hat{\varepsilon }}_{a,2}}$ is given as
\begin{equation}
{{\hat{\varepsilon }}_{a,2}}=\frac{1}{M}\sum\limits_{i=1}^{M}{{{{\hat{\varepsilon }}}_{i}}}.
\label{equ19}
\end{equation}
Now we introduce a simple strategy of parameter optimization ($\delta$, ${{k}_{a}}$ and ${{n}_{a}}$) for the proposed LCAC code to increase the cost of illegally-copying attacks. First, by considering the impact of embedding operation on the source message, the value of ${{n}_{a}}$ is determined to satisfy the covertness requirement of the proposed LCAC code. Second, through some experimental results under a practical condition, the value of $\delta$ is determined to correctly decode the authentication message for satisfying the robustness requirement of the proposed LCAC code. At last, based on a prediction function, the value of ${{k}_{a}}$ is optimized to achieve a tradeoff between the production cost and the cost of illegally-copying attacks. In the next subsection, the experimental results of the parameter optimization are presented.

\subsection{Experimental Results}
\begin{figure}[!t]
\centering
\includegraphics[width=\linewidth]{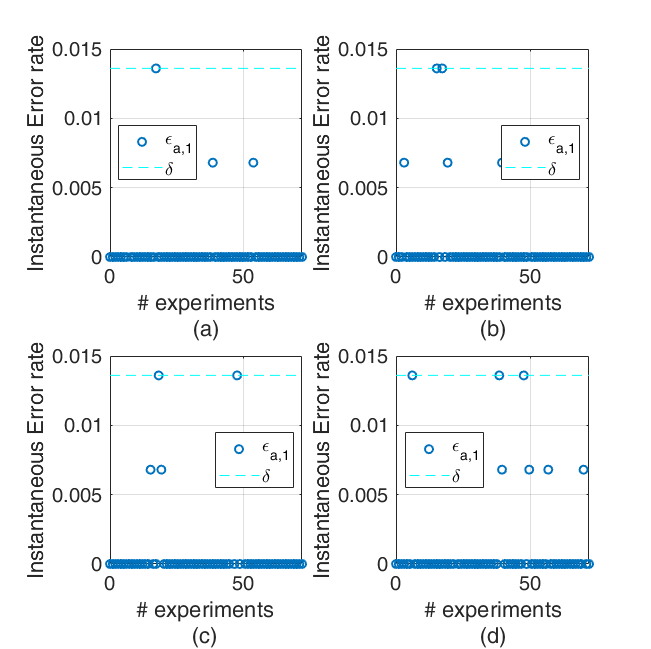}
\caption{Experimental results: instantaneous BER of authentication message for a legal 2D barcode in a combination of different printers and scanners, including (a) ${{\mathsf{P}}_{1}}-{{\mathsf{S}}_{1}}$, (b) ${{\mathsf{P}}_{1}}-{{\mathsf{S}}_{2}}$, (c) ${{\mathsf{P}}_{2}}-{{\mathsf{S}}_{1}}$, and (d) ${{\mathsf{P}}_{2}}-{{\mathsf{S}}_{2}}$, respectively. The $x$-axis represents different instances of experiments.}
\label{Figure16}
\vspace{-0.5cm}
\end{figure}

\begin{figure}[!t]
\centering
\includegraphics[height=7cm]{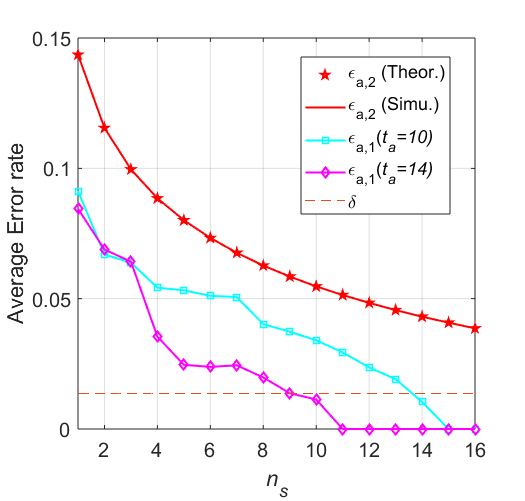}
\caption{Average BERs of authentication message for an SC attack under a mobile phone.}
\label{Figure17}
\vspace{-0.5cm}
\end{figure}

In this subsection, we present the experimental results under a mobile phone in both Fig. \ref{Figure16} and Fig. \ref{Figure17}, where the other parameters are the same as those of Tab. \ref{Table1}.
Fig. \ref{Figure16} shows the instantaneous BER of authentication message for a legal $2$D barcode, while Fig. \ref{Figure17} shows the average BER of authentication message for an SC attack.
The authentication threshold $\delta$ is also illustrated in Fig. \ref{Figure16} for a comparison purpose.
From Fig. \ref{Figure16}, we can see that some non-zero instantaneous ${{\varepsilon }_{a,1}}$ occurs occasionally although most of them are zero.
Thus, under the current conditions, we set the value of $\delta$ as the maximum instantaneous ${{\varepsilon }_{a,1}}$, \ie, $\delta =0.012$ for satisfying the robustness requirement of the proposed LCAC code.

Fig. \ref{Figure17} shows the average simulation results of ${{\hat{\varepsilon }}_{a,1}}$ and ${{\hat{\varepsilon }}_{a,2}}$ based on a prediction function, and the theoretical results of ${{\hat{\varepsilon }}_{a,2}}$ defined in (\ref{equ19}). From Fig. \ref{Figure17}, we can see that the average simulation results of ${{\hat{\varepsilon }}_{a,2}}$ matches perfectly with the corresponding theoretical results. As the estimated number of synthesized samples increases, the values of both ${{\hat{\varepsilon }}_{a,1}}$ and ${{\hat{\varepsilon }}_{a,2}}$ gradually decrease as expected. Moreover, two cases of ${{\hat{\varepsilon }}_{a,1}}$ (${{k}_{a}}=147$ and ${{k}_{a}}=179$) are simultaneously illustrated in Fig. \ref{Figure17}, which indicates that the ${{\hat{\varepsilon }}_{a,1}}$ can be controlled by setting the value of ${{k}_{a}}$. Based on this trend, the value of ${{k}_{a}}$ can be determined by making ${{\hat{\varepsilon }}_{a,1}}>\delta $, when the estimated number of synthesized samples is beyond a certain value. A simple example is given as follows to verify the efficacy of the parameter optimization.

\begin{table*}
\centering
\caption{BER comparison between before and after optimizations (Mobile Phone)}
\footnotesize
\begin{tabular}{|c|c|c|c|c|c|c|}
  \hline
  optimization & \multicolumn{3}{c|}{Before}  & \multicolumn{3}{c|}{After} \\
  \hline
  Parameters & \multicolumn{3}{c|}{${{k}_{a}}=147$, ${{t}_{a}}=14$ and ${{n}_{s}}=10$}  & \multicolumn{3}{c|}{${{k}_{a}}=179$, ${{t}_{a}}=10$ and ${{n}_{s}}=10$} \\
  \hline
  Metrics & ${{\varepsilon }_{a,2}}$ & ${{\varepsilon }_{a,1}}$ & $\mathbb{P}({{\varepsilon }_{a,1}}=0)$ & ${{\varepsilon }_{a,2}}$ & ${{\varepsilon }_{a,1}}$ & $\mathbb{P}({{\varepsilon }_{a,1}}=0)$ \\
  \hline
  Performance & 0.0342&0.0113&84.88\%&0.0471&0.0340&44.71\% \\
  \hline \hline
  Parameters & \multicolumn{3}{c|}{ } & \multicolumn{3}{c|}{${{k}_{a}}=179$, ${{t}_{a}}=10$ and ${{n}_{s}}=14$ }\\
  \cline{1-1}\cline{5-7}
  Metrics & \multicolumn{3}{c|}{N/A} & ${{\varepsilon }_{a,2}}$&${{\varepsilon }_{a,1}}$&$\mathbb{P}({{\varepsilon }_{a,1}}=0)$\\
  \cline{1-1}\cline{5-7}
  Performance & \multicolumn{3}{c|}{ } & 0.0378&0.0104&81.82\%\\
  \hline
\end{tabular}
\label{Table6}
\vspace{-0.5cm}
\end{table*}

Based on the experimental conditions of Fig. \ref{Figure16}, a BER comparison between before and after optimizations are given in Tab. \ref{Table6}, where the threshold is set as $\delta =0.012$.
From Tab. \ref{Table6}, we can see that, before optimization, the average value of ${{\varepsilon }_{a,1}}$ equals to $0.0113$ for an SC attack with ${{n}_{s}}=10$, and even $\mathbb{P}({{\varepsilon }_{a,1}}=0)$ equals to $84.88\%$.
After optimization, the value of ${{k}_{a}}$ is increased from $147$ to $179$, which indicates that the error correction capability of authentication message ${{t}_{a}}$ is reduced from $14$ to $10$.
Then, the average value of ${{\varepsilon }_{a,1}}$ is increased to $0.0340$ for an SC attack with ${{n}_{s}}=10$, and $\mathbb{P}({{\varepsilon }_{a,1}}=0)$ obviously drops to $44.71\%$.
While an attacker should increase the number of synthesized samples to ${{n}_{s}}=14$, the probability of attack success can be improved to a similar level which happens before optimization.
In other words, the results predicted in Fig. \ref{Figure17}, ${{n}_{s}}=10$ before the parameter optimization can be successfully attacked whereas ${{n}_{s}}=14$ after the parameter optimization can be successfully attacked.
Thus, an attacker should increase the number of synthesized samples to achieve the probability of attack success as a similar level which happens before optimization, as shown in Fig. \ref{Figure17}.
After supplementing the experiment, we can find that the prediction and optimization is still effective, that is, ${{n}_{s}}=10$ before the parameter optimization, ${{\varepsilon }_{a,1}}<\delta $ , ${{n}_{s}}=14$ after parameter optimization, ${{\varepsilon }_{a,1}}<\delta $.

\section{Extended Experimental Results and Discussion}
\subsection{Comparison with Two-level QR Code}
In literature, there are many existing approaches to achieve anti-copying.
We choose the latest one, which is called as Two-Level QR (2LQR) code\cite{7349185}, to compare the performance with our approach.
To the best of our knowledge, the 2LQR code is the best approach in the style of active embedding for defending against illegally-copying attacks.
The basic idea of the 2LQR code is to replace all black modules of a standard QR code with some black-and-white patterns which are unknown to the third party and increases the pixel numbers of each black module.
In comparison with the 2LQR code, our approach has the following advantages:

First, the 2LQR code introduces visually perceptual modification even if we do not put a 2D barcode with an embedded authentication message and a 2D barcode without that.
However, our approach does not have this issue if we do not put a 2D barcode with an embedded authentication message and a 2D barcode without that at the same place.
Here, we justify this conclusion by comparing the mean bias and variance of different gray values under both the 2LQR code and our approach, where the mobile phone $\mathsf{M}$ is considered as the capturing device.
Specifically, the mean bias represents the distance from the standard constellation point to the recovered constellation point.
Smaller values of the mean bias and the variance correspond to smaller visually perceptual modification.
We present the experimental results in Tab. \ref{Table7}, where we use the gray value ``0'' to represent the results of the 2LQR code since the 2LQR code only modifies the black part.
Here, we choose two different and independent patterns to replace the black modules in the 2LQR code.
Tab. \ref{Table7} includes four cases:
\begin{enumerate}

  \item Tab. \ref{Table7}(a) represents the case that the experimental results are obtained by a mobile phone, where we do not embed the authentication message in the source message;
  \item Tab. \ref{Table7}(b) represents the case that the experimental results are obtained by a mobile phone, where we embed the authentication message in the source message.
\end{enumerate}
Note that, in Tab. \ref{Table7}, the bold numbers in parenthesis represent the mean bias from the standard constellation point to the recovered constellation point, which are obtained by calculating the absolute values between standard values and experimental values.
From Tab. \ref{Table7}, we can see that the 2LQR code introduces larger both mean bias and variance than our approach.
By comparing  the results of Tab. \ref{Table7}(a) with those of Tab. \ref{Table7}(b), we can see that  the 2LQR code introduces significant difference for the mean and the variance due to embedding the authentication message, whereas our approach introduces a slight difference for both the mean and the variance.
In summary, our approach has much better covertness performance.
\begin{table}[!t]
\centering
\caption{Comparing the mean and variance of our approach with those of the 2LQR code, where we use ``0'' to represent the results of the 2LQR code and the bold numbers in parenthesis represent the mean bias from the standard constellation point to the recovered constellation point.}
\footnotesize
\begin{minipage}[t]{0.42\linewidth}
\centering
(a) Without Embedded Authentication Message
\begin{tabular}{|c|c|c|}
\hline
$x$ & $\mu \left( x \right)$ & ${{\sigma }^{2}}\left( x \right)$  \\
\hline
0	&11.45 \textbf{(11.45)}	&21.38	\\
40	&39.89 \textbf{(0.11)}	&17.76	\\
100	&99.57 \textbf{(0.43)}	&154.43	\\
160	&160.06 \textbf{(0.06)}	&129.61	\\
220	&217.77 \textbf{(2.23)}	&31.39	\\
\hline
\end{tabular}
\end{minipage}
\hspace{0.35cm}
\begin{minipage}[t]{0.42\linewidth}
\centering
(b) With Embedded Authentication Message
\begin{tabular}{|c|c|c|}
\hline
$x$ & $\mu \left( x \right)$ & ${{\sigma }^{2}}\left( x \right)$ \\
\hline
0	&69.42 \textbf{(69.42)}	&225.28	\\
40	&39.64 \textbf{(0.36)}	&23.10	\\
100	&100.75 \textbf{(0.75)}	&151.68\\
160	&160.29 \textbf{(0.29)}&143.24	\\
220	&216.61 \textbf{(3.39)}	&39.84	\\
\hline
\end{tabular}
\end{minipage}
\label{Table7}
\end{table}

Second, the 2LQR code requires higher positioning accuracy of the capturing equipment or higher proportion of the training sequence.
This is because the 2LQR code requires higher resolution to capture each sub-module whereas our approach only requires the average intensity of each entire module.
Third, our approach provides a theoretical model for illegal-copying $2$D barcode, which is verified in Section IV.
Moreover, based on the theoretical model, we optimize the parameters of our approach in order to increase the cost of copying attacks and achieve a better anti-copying effect, which is verified in Section VI.
However, the 2LQR code did not provide the above features.

\subsection{Experimental Results in Standard QR Codes}
For a standard QR code, since there are only $0$ and $255$ gray values, it has larger tolerance level to noise as compared with the case of $M=4$.
Thus, we should reduce the error-correction capability of the authentication message.
Specifically, we set ${{n}_{a}}=255$ bits, ${{k}_{a}}=247$ bits, and ${{t}_{a}}=1$ in standard QR codes.
Tab. \ref{Tab_QR} provides the experimental results of our approach in a standard QR code under both Mobile Phone and Scanner.
From Tab. \ref{Tab_QR}, we can draw the same conclusions under the case of $M=4$.
In the SPS, all decoded BERs are zeros whereas they obviously increase in the DPS, which is also used to detect whether the received $2$D
barcode is illegally copied or not.
\begin{table}[!t]
\centering
\caption{Decoded BER of our approach in a standard QR code.}
\footnotesize
\begin{tabular}{|c|c|c|c|c|c|}
  \hline
  \multirow{1}{*}{} & \multicolumn{1}{c|}{Mobile Phone} & \multicolumn{1}{c|}{Scanner} \\
  \hline
  ${{\varepsilon }_{a,1}}$ (SPS) & 0 & 0 \\
  \hline
  ${{\varepsilon }_{a,1}}$ (DPS) & 0.0064 & 0.0368  \\
  \hline
\end{tabular}
\label{Tab_QR}
\vspace{-0.5cm}
\end{table}

\subsection{Discussion}
By analyzing the above experimental results, we can draw the following conclusions:
First, our approach does not destroy the completion of $2$D barcode, since the source message in the SPS process can be successfully decoded by a legal receiver.
Second, by comparing with the experimental results based on various printers, scanners, and mobile phone, it can be found that the sample histogram and curve fitting of the theoretical model match well, so it can be concluded that the theoretical model works well.
Third, based on the theoretical model, we build a prediction function to optimize the parameters of our approach. The parameters optimization incorporates the
covertness requirement, the robustness requirement and a tradeoff between the production cost and the cost of illegally-copying attacks together.
The experimental results show that the proposed LCAC code with two printers and two scanners can detect the DC attack effectively and resist the SC attack up to the access of $14$ legal copies.

\section{Conclusion}
In this paper, the LCAC $2$D barcode was proposed, which exploited the difference between the noise characteristics of legal and illegal channels. The proposed LCAC code effectively overcomes the drawbacks of the conventional anti-copying approaches. For accurately evaluating the performance of the proposed LCAC code, we used a GGD to model a DPS process in an illegal copying attack. By comparing with the sample histogram and curve fitting of the theoretical model, the theoretical model works well. For evaluating the security of the proposed LCAC code, besides the DC attack, the improved version which is the SC attack was also considered in this paper. We built a prediction function to optimize the parameters of the proposed LCAC code based on the theoretical model. The parameters optimization incorporated the covertness requirement, the robustness requirement and a tradeoff between the production cost and the cost of illegally-copying attacks together. The experimental results showed that the proposed LCAC code is able to prevent illegal copying effectively.

There are several promising future directions based on the proposed LCAC $2$D barcode.
First, it is natural to extend the gray-scale barcodes in this work to color barcodes.
Although the color barcodes are more complicated than the gray-scale barcodes, both barcodes are quite similar and thus LCAC $2$D color barcodes are likely able to prevent illegally copying effectively.
Second, we research on detection techniques to evaluate the security level of various anti-copying $2$D barcodes.
At last, we use some approaches of machine learning or deep learning to improve the performance of the proposed LCAC $2$D barcode.

\appendices
\numberwithin{equation}{section}

\section{Analysis of the Two Embedding Strategies}


%
We consider two scenarios in practical applications of $2$D barcodes: in the first scenario, there is no occlusion to emulate an ideal situation, while in the second scenario, there is a small occlusion over a $2$D barcode to emulate a non-ideal situation.
Three examples of a small occlusion are considered and the experimental results are given in Tab. \ref{Table_Strategy}.
The size of occlusion is defined as $a\times b$, where $a$ and $b$ are the numbers of row occlusion and column occlusion, respectively.
Note that the case of $0\times 0$ represents the first scenario, in which there is no occlusion.


\begin{table*}[!t]
\centering
\caption{Robustness analysis of two embedding strategies.}
\footnotesize
\begin{tabular}{|c|c|c|c|c|c|c|c|c|}
  \hline
  \multirow{2}{*}{Size (in modules)} & \multicolumn{4}{c|}{Strategy $1$} & \multicolumn{4}{c|}{Strategy $2$} \\
  \cline{2-9}
  &${{\varepsilon }_{c,2}}$	& ${{\varepsilon }_{c,1}}$ & ${{\varepsilon }_{a,2}}$ & ${{\varepsilon }_{a,1}}$ & ${{\varepsilon }_{c,2}}$ & ${{\varepsilon }_{c,1}}$ & ${{\varepsilon }_{a,2}}$ & ${{\varepsilon }_{a,1}}$ \\
  \hline
  $0\times 0$ & 0.0313 & 0 & 0 & 0 & 0.0314 & 0 & 0 & 0\\
  \hline
  $2\times 11$ & 0.0519 & 0.0073 & 0.0230 & 0.0179 & 0.0521 & 0.0073 & 0.0211 & 0.0100\\
  \hline
  $4\times 11$ & 0.0566 & 0.0074 & 0.0256 & 0.0185 & 0.0567 & 0.0079 & 0.0247 & 0.0104\\
  \hline
  $7\times 11$ & 0.0724	& 0.0147 & 0.044 & 0.0339 & 0.0731 & 0.0161	& 0.0417
   & 0.0224\\
  \hline
\end{tabular}
\label{Table_Strategy}
\vspace{-0.5cm}
\end{table*}

For the first scenario, we can see that all BERs are zeros except ${{\varepsilon }_{c,2}}$. This is because the embedding of authentication message sacrifices the robustness of the source message; however, the error-correction capabilities of encoding modules for both source and authentication messages are sufficiently powerful and channel distortion is not introduced, we obtain ${{\varepsilon }_{c,1}}={{\varepsilon }_{a,2}}={{\varepsilon }_{a,1}}=0$. Moreover, we observe that ${{\varepsilon }_{c,2}}$ in Strategy $1$ is smaller than that in Strategy $2$, which reflects that Strategy $1$ has better covertness performance than that of Strategy $2$. It should be noted that this result verifies Observation $1$ given in Section III.

For the second scenario, we can see that a larger occlusion size leads to higher BER values in all metrics. Moreover, two additional conclusions are drawn here. First, since the values of ${{\varepsilon }_{c,1}}$ and ${{\varepsilon }_{c,2}}$ in Strategy $1$ are larger than those in Strategy $2$, which is consistent with the first scenario, Strategy $1$ has better performance in covertness than Strategy $2$ and Observation $1$ is verified again. Second, since the values of ${{\varepsilon }_{a,1}}$ and ${{\varepsilon }_{a,2}}$ in Strategy $2$ are smaller than those in Strategy $1$, Strategy $2$ has better robustness performance than that of Strategy $1$ and Observation $2$ is verified.

\bibliographystyle{IEEEtran}
\bibliography{IEEEabrv,paper}
\end{document}